\documentclass{aa}  

\usepackage{amssymb}
\usepackage{graphicx}
\usepackage[utf8]{inputenc}
\usepackage{amsmath}
\usepackage{lineno}
\usepackage[squaren]{SIunits}
\usepackage{comment}
\usepackage{nicefrac}
\usepackage[shortlabels]{enumitem}
\usepackage{placeins}
 \usepackage[version=4]{mhchem}
 \usepackage{hyperref}
\hypersetup{colorlinks=true, linkcolor=blue, breaklinks=true,  urlcolor= blue, citecolor=blue} 
\usepackage[varg]{txfonts}

\begin{document} 

\title{The surface of (1) Ceres in visible light as seen by Dawn/VIR\thanks{The HipS files are available at the CDS via \url{http://alasky.
u-strasbg.fr/pub/10.1051_0004-6361_202038512}}}
\subtitle{Reproduced with permission from Astronomy \& Astrophysics, ©ESO}

\author{B. Rousseau \inst{1}
  \and M.C. De Sanctis \inst{1}
  \and A. Raponi \inst{1}
  \and M. Ciarniello \inst{1}
  \and E. Ammannito \inst{2}
  \and A. Frigeri \inst{1}
  \and M. Ferrari \inst{1}
  \and S. De Angelis \inst{1}
  \and F.C. Carrozzo \inst{1}
  \and F. Tosi \inst{1}
  \and S. E. Schröder \inst{3}
  \and C. A. Raymond \inst{4}
  \and C. T. Russell \inst{5}
  }

   \institute{Istituto Nazionale di Astrofisica (INAF) - Istituto di Astrofisica e Planetologia Spaziali (IAPS), Via Fosso del Cavaliere, 100, 00133, Rome, Italy\\
              \email{batiste.rousseau@inaf.it}
         \and
             Italian Space Agency (ASI), Via del Politecnico, 00133, Rome, Italy
         \and
             Deutsches Zentrum für Luft- und Raumfahrt (DLR), 12489 Berlin, Germany
         \and
             Jet Propulsion Laboratory, California Institute of Technology, Pasadena, USA
         \and
             University of California Los Angeles, Earth Planetary and Space Sciences, Los Angeles, CA, USA
            }

   \date{Received 27 May 2020; accepted 29 July 2020}

 
  \abstract
   {}
   {We study the surface of Ceres at visible wavelengths, as observed by the Visible and InfraRed mapping spectrometer (VIR)  onboard the Dawn spacecraft, and analyze the variations of various spectral parameters across the whole surface. We also focus on several noteworthy areas of the surface of this dwarf planet.}
   {We made use of the newly corrected VIR visible data to build global maps of a calibrated radiance factor at \unit{550}{\nano\meter}, with two color composites and three spectral slopes between \unit{400}{\nano\meter} and \unit{950}{\nano\meter}. We have made these maps available for the community via the Aladin Desktop software.}
   {Ceres' surface shows diverse spectral behaviors in the visible range. The color composite and the spectral slope between \unit{480}{\nano\meter} and \unit{800}{\nano\meter} highlight fresh impact craters and young geologic formations of endogenous origin, which appear bluer than the rest of the surface. The steep slope before \unit{465}{\nano\meter} displays very distinct variations and may be a proxy for the absorptions caused by the \mbox{\ce{O2- -> Fe3+}} or the \ce{2Fe^{3+} -> Fe^{2+} + Fe^{4+}} charge transfers, if the latter are found to be responsible for the drop in this spectral range. We notice several similarities between the spectral slopes and the abundance of phyllosilicates detected in the infrared by the VIR, whereas no correlation can be clearly established with carbonate species. The region of the Dantu impact crater presents a peculiar spectral behavior -- especially through the color and the spectral slope before \unit{465}{\nano\meter} -- suggesting a change in composition or in the surface physical properties that is not observed elsewhere on Ceres.}
   {}
   
   
   \maketitle
%
\section{Introduction}
\label{Sec_Introduction}

The NASA Dawn spacecraft was launched on September 27, 2007, reaching the asteroid Vesta in July 2011 \citep{2007_Russell}. The spacecraft then left Vesta in September 2012 and entered orbit around Ceres in March 2015. The study of the dwarf planet Ceres lasted more than three years, up until the end of October 2018.\par
Dawn was equipped with three instruments to study the surfaces of Vesta and Ceres: a Framing Camera (FC), with one clear filter and seven narrow band-pass filters \citep{2011_Sierks}, which provided optical imagery in the visual range at a high spatial resolution (up to \unit{30}{\meter}/pixel during the low-altitude orbit phase at Ceres); a Gamma Ray and Neutron Detector (GRaND), which sampled the elemental composition of the surface at coarser spatial scales \citep{2011_Prettyman}; and a Visible InfraRed mapping spectrometer (VIR), a hyperspectral imager combining spectroscopic and imaging capabilities in the visible to infrared wavelengths \citep{2011_De_Sanctis}. In this work, we focus on Dawn's observations of Ceres  at visible wavelengths as measured by the Visible and InfraRed mapping spectrometer (VIR). \par
The Dawn orbital mission at Ceres was split into different ranges of altitude over the surface \citep{2007_Russell, 2018_De_Sanctis_a}. For details on the different phases of the mission, specifically for the visible channel, see \cite{2019_Rousseau_c}. This allowed for observations of the surface at moderate and high spatial resolutions, achieving a nearly global mapping and providing an unprecedented view of the surface of Ceres.\par
Overall, the surface of Ceres is characterized by an almost flat reflectance spectrum in the spectral region below \unit{2.6}{\micro\meter}, with the exception of a broad inflection at about \unit{1.2}{\micro\meter}, possibly due to iron-bearing materials \citep{2011_Rivkin, 2018_De_Sanctis_a}. The \unit{2.6-4.2}{\micro\meter} wavelength region is characterized by a broad asymmetric feature, in which several distinct absorptions show up at $2.72$, $3.05-3.1$, $3.3-3.5$, and \unit{3.95}{\micro\meter}. The strongest, a narrow feature centered at \unit{2.72-2.73}{\micro\meter}, is indicative of the structural \ce{OH} in \ce{Mg}-bearing phyllosilicates. The other signatures are attributed to carbonates ($\sim$\unit{3.9}{\micro\meter}) and ammoniated phyllosilicates ($\sim$\unit{3.06}{\micro\meter}). The spectra show a broad absorption between \unit{3.3-3.6}{\micro\meter}, which may be the result of overlapping bands due to carbonates, ammoniated-phyllosilicates, and organics \citep{1998_Moroz,2002_Beran, 2008_Bishop}.\par
The average Ceres spectrum has been interpreted as the result of a mixture of dark material, \ce{Mg}-phyllosilicates, ammoniated–phyllosilicates and (\ce{Mg}, \ce{Ca})-carbonates \citep{2015_de_Sanctis_b, 2018_De_Sanctis_a}. These materials are present everywhere: no large km-sized areas lacking the above-mentioned species have been observed \citep{2016_Ammannito, 2018_Carrozzo}. However, changes in the strength of the absorption features have been reported \citep{2016_Ammannito, 2018_Carrozzo}, indicating variability in the relative abundance of the phyllosilicates and carbonates. Similarly, ammonium-bearing minerals are ubiquitous on the surface of Ceres, even though they show a difference in abundance and in chemical form \citep{2016_De_Sanctis, 2016_Ammannito}. The observed mineralogy requires pervasive and long-standing aqueous alteration \citep{2016_De_Sanctis, 2016_Ammannito}, as also suggested by the spatial uniformity of element abundance measurements of equatorial regolith \citep{2016_Prettyman, 2018_Lawrence}.\par
The global map of hydrogen abundance obtained by GRaND indicates the presence of ice buried below the Cerean regolith, which is increasingly abundant moving away from the equator to high latitudes \citep{2016_Prettyman, 2018_Lawrence}. Indeed, local exposure of water ice has been identified on the surface of Ceres, for example, in the craters Oxo and Juling \citep{2016_Combe, 2018_Raponi, 2019_Combe_a}. Sodium carbonates were first detected in the crater Occator's bright faculae \citep{2016_De_Sanctis} and later identified in many other bright areas \citep{2017_Zambon, 2018_Carrozzo}. Moreover, regional areas with organic material have also been detected (\cite{2017_De_Sanctis, 2018_De_Sanctis_b}).\par
So far, the spectral behavior of Ceres' surface at visible wavelengths, as measured by VIR, has not been deeply investigated due to the presence of instrumental artifacts that have only recently been corrected by \cite{2019_Rousseau_c}. Taking advantage of this new spectral calibration, in this work, we provide a global analysis of VIR visible spectral properties and provide updated maps of visible reflectance and related spectral parameter maps.\par
Details about the VIR instrument, the dataset, and the definition of the spectral parameters, along with their mapping are given in Sect. \ref{Sec_Data and methods}. In Sect. \ref{Sec_Maps of the spectral parameters}, we describe the global maps of the spectral parameters. In Sect. \ref{Sec_Area_of_interest} and \ref{Sec_General_discussion} we discuss our results in  light of the existing literature, as well as the VIR infrared observations and of the Framing Camera results. Finally, we present our summary in Sect. \ref{Sec_Conclusion}.

\section{Data and methods}
\label{Sec_Data and methods}

\subsection{The VIR instrument}
\label{SubSec_The VIR instrument}
The VIR instrument is an imaging spectrometer that combines spatial and spectral information. It is made up of two channels, the first working in the visual wavelengths (VIS, \unit{0.25}{\micro\meter} - \unit{1.07}{\micro\meter}) and the second in the infrared (IR, \unit{1.02}{\micro\meter} - \unit{5.09}{\micro\meter}). The average spectral sampling in the VIS is 1.8 nm/band and its instantaneous field of view (IFOV) of \unit{250}{\micro\radian} $\times$ \unit{250}{\micro\radian}. A detailed description of the instrument is provided by \cite{2011_De_Sanctis}. Here, we focus on the visible dataset acquired by VIR at Ceres, but we limit the analysis in the spectral interval between \unit{400}{\nano\meter} and \unit{950}{\nano\meter} because of low signal conditions combined with calibration residuals outside this range.

\subsection{Data correction and processing}
\label{SubSec_Data correction and processing}
The data used in this study correspond to the calibrated version of the VIR observations (LEVEL 1B) available through the Planetary Data System (PDS) online data archive\footnote{\url{https://sbn.psi.edu/pds/resource/dawn/dwncvirL1.html}}. They are expressed in units of calibrated radiance factor, as described by \cite{2016_Carrozzo}. Several corrections have been applied over this calibrated version, which are briefly described below.\par
A multiplicative matrix is applied on the data in order to correct for the odd-even effect, spectral spikes, vertical stripes, and systematic artifacts. This first procedure is described in \cite{2016_Carrozzo}. The VIR VIS spectra are affected by a positive slope which has been corrected a first time by \cite{2016_Carrozzo}, using ground-based observations. Here we replaced and refined this correction by means of the following steps: 1) we collected ground observations of Ceres, which are mutually consistent in the spectral range where they overlap \citep{1979_Chapman_a, 1994_Roettger, 2002_Parker, 2002_Bus_b, 2002_Bus_a, 2006_Lazzaro, 2006_Li, 2011_Rivkin}; 2) then each ground full-disk observation (point n$\degr1$) was converted in bidirectional reflectance at standard viewing geometry ($\text{incidence angle}=30\degr$, $\text{emission angle}=0\degr$, $\text{phase angle}=30\degr$) by means of Hapke modeling \citep{2012_Hapke}, according to the photometric parameters derived by \cite{2017_Ciarniello};
3) based on the ground-based spectra (point n$\degr2$), we calculated a smooth average spectrum which covers the whole spectral range of the visible channel of the VIR spectrometer; 4) we collected VIR data at standard viewing geometry ($\text{incidence angle}=30\degr$, $\text{emission angle}=0\degr$, $\text{phase angle}=30\degr$) and we calculated the average spectrum; 5) finally, we calculated the ratio between the average spectrum from ground observations (point n$\degr3$) and the average spectrum obtained from VIR data (point n$\degr4$). This ratio spectrum is used as a multiplicative correction factor for every single VIR spectrum.\par
The data were then photometrically corrected, as described by \cite{2017_Ciarniello}. This allows the variability of the observation geometry to be discarded. Finally, to overcome spurious spectral variations due to the detector temperature, an empirical correction was developed, as detailed in \cite{2019_Rousseau_c}, and applied to the data used in this study.\par
The data investigated in this work were acquired over four different phases of the Dawn mission at Ceres, lasting from late April 2015 to mid-August 2015 (see Table \ref{Table1} for more details). The dataset is made up of 505 hyperspectral cubes, corresponding to nearly 8 million single observations. It provides an almost complete coverage of Ceres' surface and a high redundancy, particularly in the $50\degr$S--$50\degr$N latitude range (see Fig. \ref{Fig_Appendix_DENSITY_MAP} in Appendix \ref{Appendix_Density_maps}).\par
To deal with this large amount of data, we developed an automatic procedure aimed at calculating different spectral parameters of interest for all the observations of each cube contained in a single mission phase. The results are stored in binary FITS tables to facilitate further processing.
We did not apply any filters on the observation angles; this allows us to take advantage of the full dataset. The counterpart is a lower quality rendering at high latitudes due to a less accurate photometric correction for more extreme observation angles \citep{2017_Ciarniello}. Finally, in each map, we filtered out several cubes that showed too many artifacts (see Sect. \ref{Sec_Maps of the spectral parameters}).
\begin{table}

\caption{Mission phases of Dawn at Ceres used in this study.}
\label{Table1}
\centering
\begin{tabular}{c c c c c}
\hline \hline
\begin{tabular}{@{}l@{}}Mission\\ Phase\end{tabular} & \begin{tabular}{@{}l@{}}Start date\\(mm-dd)\end{tabular} & \begin{tabular}{@{}l@{}}Stop date\\(mm-dd)\end{tabular} & \begin{tabular}{@{}l@{}}Cubes\\(used/total)\end{tabular} & \begin{tabular}{@{}l@{}}Resolution\\ (m/pix)\end{tabular} \\[5pt]
\hline
CSR & 04-25 & 05-07 & 75/75 & 3400--3500
\\[2.5pt]
CTS & 05-16 & 05-22 & 12/12 & 1300--1800
\\[2.5pt]
CSS & 06-05 & 06-27 & 230/280 & 1000--1100
\\[2.5pt]
CSH & 08-18 & 10-21 & 188/196 & 360--400
\\ \hline
\end{tabular}
\tablefoot{Mission phases are chronologically sorted and we report only the periods during which VIR visible data were acquired. CSR: Ceres Science Rotational Characterization
3; CTS: Ceres Transfer to Survey; CSS: Ceres Science Survey; CSH: Ceres Science High Altitude Mapping Orbit Ceres. These data were all acquired in 2015. In the fourth column, a discrepancy between processed versus available data arises from the occurrence of sky observation or corrupted data. The fifth column provides the approximate minimum and maximum across-track resolutions.}
\end{table}
%
\subsection{Spectral parameters}
\label{SubSec_Spectral parameters}
\begin{figure}
    \resizebox{\hsize}{!}
    {\includegraphics{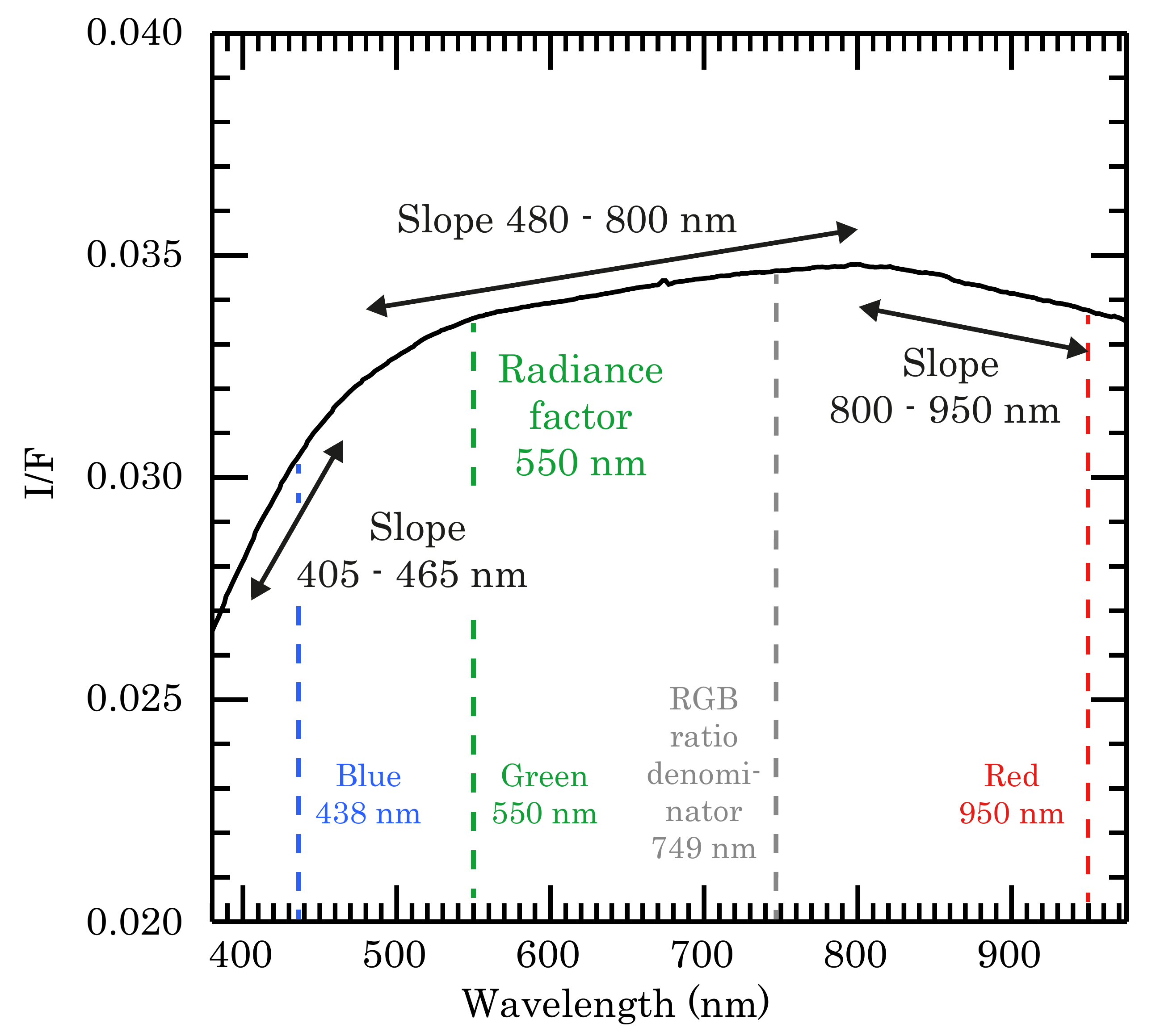}}
    \caption{\label{Fig_MED_SPEC} Median spectrum of Ceres' surface with the spectral parameters used in this study. RGB channels used to build the RGB composite are indicated at the bottom.}
\end{figure}

The spectral variability at the surface of Ceres in the \unit{400-950} {\nano\meter} range was studied through several parameters based on the average spectrum as shown in Fig. \ref{Fig_MED_SPEC}. This spectrum corresponds to the median of the CSH mission phase observations. Due to their spatial distribution, they are representative of the whole Ceres surface (see Fig. \ref{Appendix_Density_maps} of the Appendix and \cite{2019_Rousseau_c}); the resulting spectrum can be considered as a "mean Ceres" and reliable for the spectral parameter definition.\par
The first spectral parameter is the calibrated radiance factor at \unit{550}{\nano\meter} -- hereafter called reflectance \footnote{The reflectance and the radiance factor are proportional by a factor of $\pi$ \citep{2012_Hapke}.} or $I/F_{550nm}${}. Then, we defined two RGB color composites chosen to closely resemble the ones adopted by \cite{2017_Schroder} and \cite{2016_Nathues}. The first is a simple RGB composite, based on the red, green, and blue colors attributed to the reflectance at \unit{950}{\nano\meter}, \unit{550}{\nano\meter}, and \unit{438}{\nano\meter}, respectively, which shows color and albedo variations. This is both an advantage and an inconvenience since the perception of the albedo variations affects the perception of the colors and vice versa. The second is a RGB ratio -- based on the red, green, and blue colors attributed to the reflectance ratios, $\nicefrac{I/F_{950nm}}{I/F_{749nm}}$, $\nicefrac{I/F_{550nm}}{I/F_{749nm}}$ and $\nicefrac{I/F_{438nm}}{I/F_{749nm}}$ , respectively. This allows us to get rid of the albedo surface changes and emphasize only the color variations. These two RGB composites are, thus, complementary and useful in the visible spectral domain.\par
Finally, we used three spectral slopes to enhance the relative spectral variations across the surface. Following the definition adopted by \cite{2015_Ciarniello, 2017_Ciarniello}, each slope, expressed in \reciprocal{\kilo\angstrom}, is defined by the Equation \ref{eqn:SPECTRAL_SLOPE} and written as $S_{\lambda_1-\lambda_2}$ , where $\lambda_1$ and $\lambda_2$ are the wavelengths of each side.
\begin{equation}
\label{eqn:SPECTRAL_SLOPE}
        S_{\lambda_1\text{-}\lambda_2}=
        \frac{(\nicefrac{I}{F})_{\lambda_2}-(\nicefrac{I}{F})_{\lambda_1}}
        {(\nicefrac{I}{F})_{\lambda_1}\times(\lambda_2-\lambda_1)}
.\end{equation}
The slope, $S_{405-465nm}$ , characterizes the steep drop of reflectance observed before $\sim$\unit{480}{\nano\meter}. The second slope, $S_{480-800nm}$ , describes the central part of the spectra, which is mostly flat and between \unit{480}{\nano\meter} and \unit{800}{\nano\meter}. The third slope is $S_{800-950nm}$: this range is chosen to characterize the near-IR part of the channel, where a slight negative slope is observed.\par 

%

\subsection{Map projections}
\label{SubSec_Map projections}
The maps presented in Sect. \ref{Sec_Maps of the spectral parameters} are built in two steps. First, the binary tables (see Sect. \ref{SubSec_Data correction and processing}) are fed into the \mbox{TOPCAT} software (\cite{2005_Taylor}) and the maps are generated using the Plate Carée projection. In those maps, each observation is represented as a point and the median is calculated in cases of overlapping. While the representation of an observation by a point is not as precise as the projection of the real footprint at the surface, a quality result is guaranteed, thanks to the scale of the map and the resolution of the data, as well as their available density \citep{2019_Rousseau_a}.\par
The second step consists of the construction of a Hierarchical Progressive Survey (HiPS) with the Hipsgen program of the Centre de Données Astronomiques de Strasbourg (CDS). The HiPS is a multi-resolution data structure \citep{2015_Fernique}, based on the Hierarchical Equal Area isoLatitude Pixelization (HEALPix). The HiPS is generated using the TOPCAT maps as inputs and the results are visualized in the Aladin Desktop software\footnote{\url{https://aladin.u-strasbg.fr/AladinDesktop/}} \citep{2000_Bonnarel}, also developed by the CDS.\par
The maps of this study are exported from Aladin Desktop in a Mollweide projection. The $I/F_{550nm}$ and RGB maps (Figs. \ref{Fig_IoF_550}, \ref{Fig_RGB}, and \ref{Fig_RGB_ratio}) do not benefit from a Framing Camera context map to avoid any misinterpretation. On the contrary, the three spectral slopes maps we present (in Sect. \ref{Sec_Maps of the spectral parameters}, Figs. \ref{Fig_SLOPE_405_465}, \ref{Fig_SLOPE_480_800}, and \ref{Fig_SLOPE_800_950}) are superimposed to a transparency of $40\%$ on a Framing Camera map. The FC map used as a background is a HiPS reprocessed in Aladin Desktop from a FC clear filter map of \cite{2016_Roatsch_c} made of data acquired during the Low Altitude Mapping Orbit (LAMO) mission phase (see Fig. \ref{Fig_Appendix_LAMO_MAP} in Appendix). This allows us to benefit from a surface context at high spatial resolution, as provided by the FC. These three maps are also presented in the appendix without the FC context and coordinate grid (Figs. \ref{Fig_APPENDIX_SLOPE_405_465_NoTransp_NoGrid}, \ref{Fig_APPENDIX_SLOPE_480_800_NoTransp_NoGrid}, and \ref{Fig_APPENDIX_SLOPE_800_950_NoTransp_NoGrid} in Appendix). The FC LAMO map and the VIR $I/F_{550nm}$, RGB composites and spectral slope maps are also freely available through the Aladin Desktop software.\par
We added masks above $75\degr$N and below $60\degr$S on the maps presented in Sect. \ref{Sec_Maps of the spectral parameters} because of a lack of data and a lower efficiency on the part of the photometric correction in order to avoid any misinterpretation. Finally, features of the Ceres surface which are discussed in the text (craters or remarkable structures) are indicated by the numbers on the maps. These numbers refer to Table \ref{Table2}, which provides the name and the coordinates for each.\par

\begin{table}
\caption{Main features mentioned in the text.}
\label{Table2}
\centering
\begin{tabular}{c c c c}
\hline \hline
\begin{tabular}{@{}l@{}}\#\end{tabular} &
\begin{tabular}{@{}l@{}}Ceres surface\\formation names\end{tabular} & \begin{tabular}{@{}l@{}}Longitude\end{tabular} & \begin{tabular}{@{}l@{}}Latitude\end{tabular}\\[5pt]
\hline
1 & Haulani & $310\degr$E & $0\degr$N
\\[2.5pt]
2 & Ikapati & $43\degr$E & $31\degr$N
\\[2.5pt]
3 & Ernutet & $44\degr$E & $52\degr$N
\\[2.5pt]
4 & Omonga & $71\degr$E & $58\degr$N
\\[2.5pt]
5 & Gaue & $86\degr$E & $31\degr$N
\\[2.5pt]
6 & Braciaca & $84\degr$E & $23\degr$S
\\[2.5pt]
7 & Kerwan & $125\degr$E & $10\degr$S
\\[2.5pt]
\ldots & Vendimia Planitia\tablefootmark{a} & $85-185\degr$E & $65\degr$N-$20\degr$S
\\[2.5pt]
8 & Dantu & $138\degr$E & $25\degr$N
\\[2.5pt]
9 & Centeotl & $141\degr$E & $20\degr$N
\\[2.5pt]
10 & Cacaguat & $143.5\degr$E & $1\degr$S
\\[2.5pt]
11 & Juling & $168\degr$E & $36\degr$S
\\[2.5pt]
12 & Kupalo & $173\degr$E & $39\degr$S
\\[2.5pt]
13 & Nawish & $192\degr$E & $18\degr$N
\\[2.5pt]
14 & Consus & $200\degr$E & $20\degr$S
\\[2.5pt]
15 & Azacca & $217\degr$E & $7\degr$S
\\[2.5pt]
16 & Lociyo & $228\degr$E & $6\degr$N
\\[2.5pt]
\ldots & Hanami Planum\tablefootmark{b} & $200-260\degr$E & $40\degr$N-$15\degr$S
\\[2.5pt]
17 & Occator & $240\degr$E & $20\degr$N
\\[2.5pt]
18 & Tawals & $237\degr$E & $38\degr$S
\\[2.5pt]
19 & Urvara & $247\degr$E & $45\degr$S
\\[2.5pt]
20 & Nunghui & $271\degr$E & $54\degr$S
\\[2.5pt]
21 & Yalode & $293\degr$E & $40\degr$S
\\[2.5pt]
22 & Fejokoo & $312\degr$E & $30\degr$N
\\[2.5pt]
23 & Xevioso & $311\degr$E & $0.6\degr$ 
\\[2.5pt]
24 & Ahuna Mons\tablefootmark{c} & $316\degr$E & $10\degr$ 
\\[2.5pt]
25 & Oxo\tablefootmark{d} & $0\degr$E & $42\degr$N 
\\[2.5pt]

\\ \hline
\end{tabular}
\tablefoot{The first column corresponds to the identification number present on the global maps in Sect. \ref{Sec_Maps of the spectral parameters}. They are ordered from west to east and from north to south. The second column is the name of the formation, while the last two correspond to the longitude and latitude coordinates. All the formations are impact craters, except for Vendimia Planitia, Hanami Planum, and Ahuna Mons.\\
\tablefoottext{a}{Vendimia region is a planitia, i.e. a plain of low altitudes interpreted to be an ancient crater basin \citep{2016_Marchi}. It includes the Dantu and Kerwan craters. Coordinates are from \cite{2018_Stephan, 2016_Preusker,2016_Roatsch_b}. See also Fig. \ref{Fig_Appendix_LAMO_MAP} in Appendix.}
\tablefoottext{b}{Hanami region is a planum, i.e. a plateau of high altitudes. It includes the Occator crater. Coordinates are derived from \cite{2018_Buczkowski}. See also Fig. \ref{Fig_Appendix_LAMO_MAP} in Appendix.}
\tablefoottext{c}{Ahuna Mons is the highest mountain on Ceres and very likely cryovolcanic in origin \citep{2016_Ruesch}.}
\tablefoottext{d}{Oxo is not completely visible on the maps in Sect. \ref{Sec_Maps of the spectral parameters}, due to the Mollweide projection and its location on the meridian of origin. Oxo is seen in Fig. \ref{Fig_Oxo_Zoom} in Sect. \ref{subSec_Oxo}}.
}
\end{table}
%
%
%
\section{Global maps of the spectral parameters}
\label{Sec_Maps of the spectral parameters}
\subsection{Map of reflectance at \unit{550}{\nano\meter}}
\label{SubSec_IoF 550nm map}
\begin{figure*}
    \centering
    \includegraphics[width=17cm]{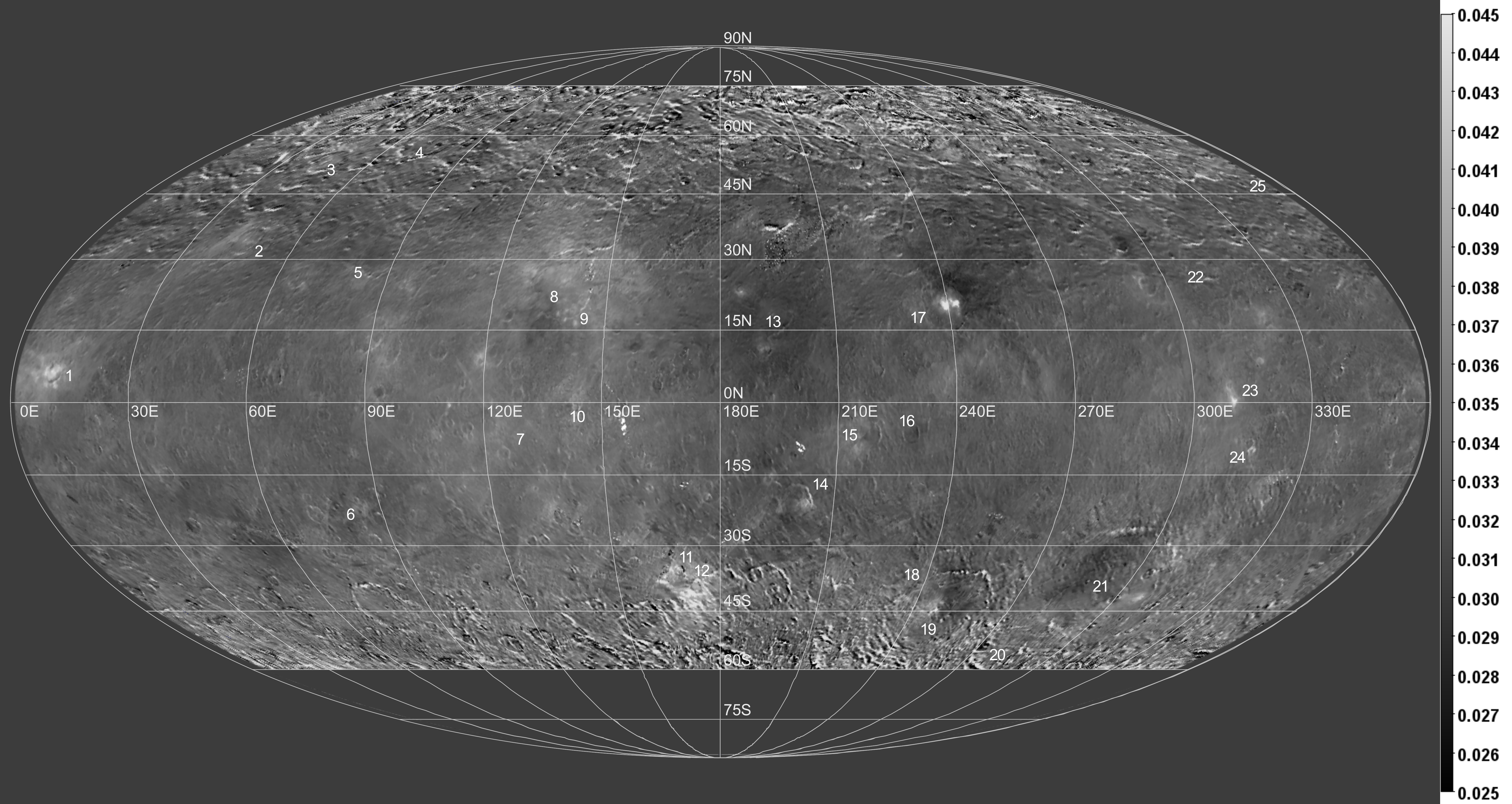}
    \caption{\label{Fig_IoF_550} Map of the VIR reflectance at \unit{550}{\nano\meter}. Numbers refer to the features of Table \ref{Table2}, discussed in the text. White areas correspond to missing data or overexposed spots (e.g., Occator faculae, n\degr17).}
\end{figure*}

Figure \ref{Fig_IoF_550} represents the map of the reflectance at \unit{550}{\nano\meter,}  highlighting its variability across the surface of Ceres. We excluded a total of eight cubes that produced major artifacts. The scale, between 0.025 and 0.045, and the filtering include 94$\%$ of the initial 8 million single observations. Some artifacts are still visible in this map and can be of different origins: 1) the effect of the shadows is noticeable at high latitudes, where the photometric correction fails to correct those areas where the geometry of observations is extreme and the signal is very low; 2) possible residuals in the corrections applied to the data (e.g., for the sensor temperature effects, see Sect. \ref{SubSec_Data correction and processing}; such kind of artifact is visible as a stripe at, e.g., west of the Braciaca crater); and 3) some small artifacts due to the projection process used to build the maps (e.g. at $195-215\degr$E -- $30-35\degr$N).\par
Nonetheless, we are able to distinguish a certain level of heterogeneities across the surface, which is characterized by median value $I/F_{550nm}$ of 0.034. This is the case for the center of Vendimia Planitia, which presents an average reflectance of 0.035. Vendimia Planitia owns two major impact craters, Kerwan and Dantu, of which Dantu is easily recognizable because of the contrast in reflectance between the south of the crater and its ejecta which are darker (mean $I/F_{550nm}$ of 0.034) than the northern part (mean $I/F_{550nm}$ of 0.037).\par
Out of Vendimia Planitia, we note different locations with a reflectance that is lower than the average surface. This is the case of the Nawish crater and its ejecta, as well as the internal northeastern and northwestern floors of the Urvara and Yalode craters, and around Occator (for more, see Sect. \ref{subSec_Occator}).\par
Ceres is known to have bright spots on its surface, that is, bright material units that stand out with respect to surrounding terrains \citep{2019_Stein}. On the reflectance map, the most distinctive bright spots are the two Cerealia and Vinalia faculae within the Occator crater, the young Haulani crater, the region around Kupalo and the Oxo (not visible due to the projection and its location at $0\degr$E; for details on Oxo, see \ref{subSec_Oxo}) and Xevioso craters. Juling and Kupalo are sufficiently visible in Fig. \ref{Fig_IoF_550}; they actually show high reflectance, but some artifacts, due to the unfavorable observation geometries occurring in those high-latitude areas, are  still present around them, particularly to the south of the craters.\par

%

\subsection{Color composite maps}
\label{SubSec_Color composite maps}

\begin{figure*}
    \centering
    \includegraphics[width=17cm]{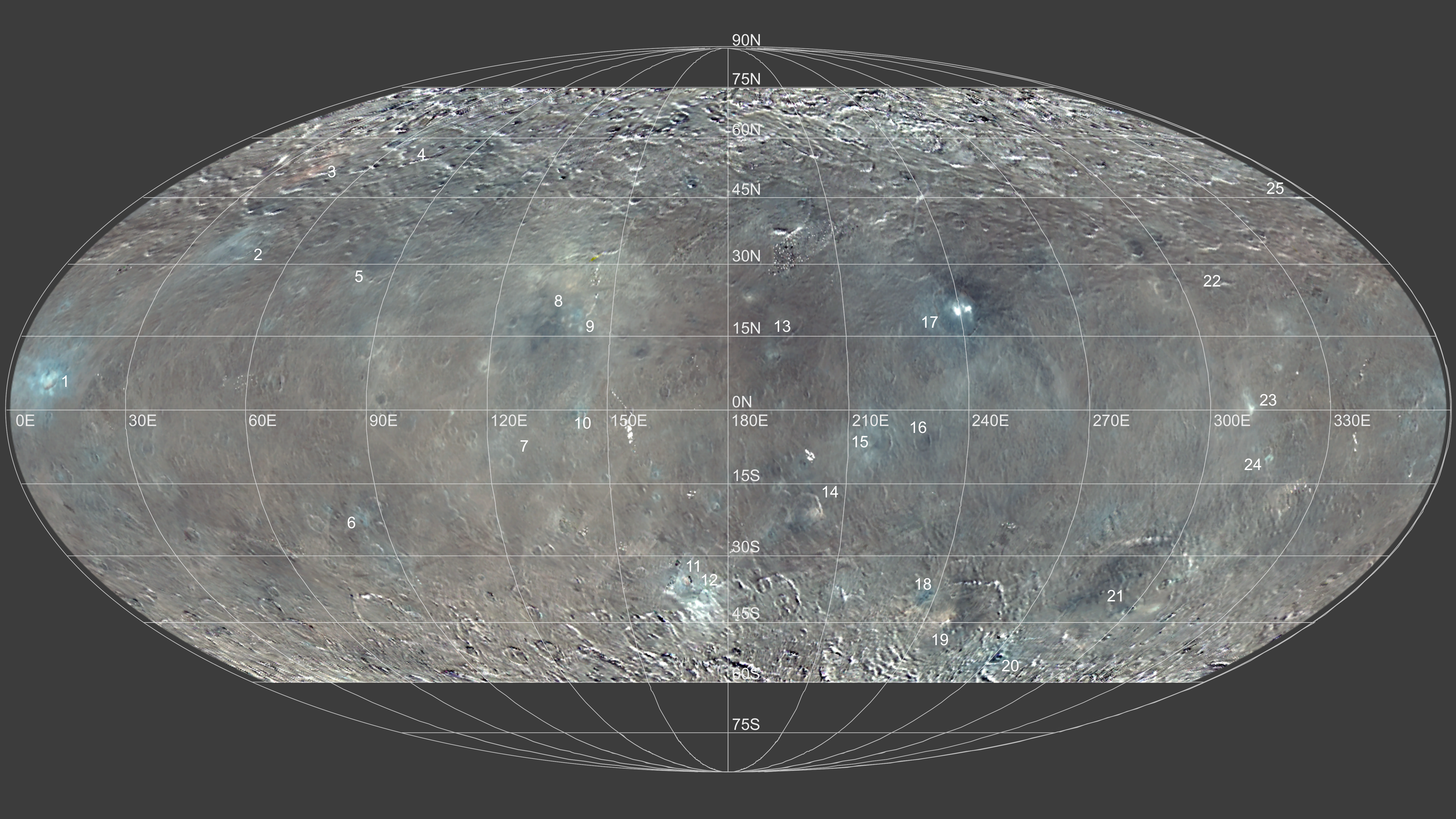}
    \caption{\label{Fig_RGB} VIR color composite map made using \unit{950}{\nano\meter}, \unit{550}{\nano\meter}, and \unit{438}{\nano\meter} for the red, green, and blue channels respectively. Numbers refer to the features of Table \ref{Table2}, as discussed in the text.  White areas correspond to missing data or overexposed spots (e.g., Occator faculae, n\degr17).}
\end{figure*}

\begin{figure*}
    \centering
    \includegraphics[width=17cm]{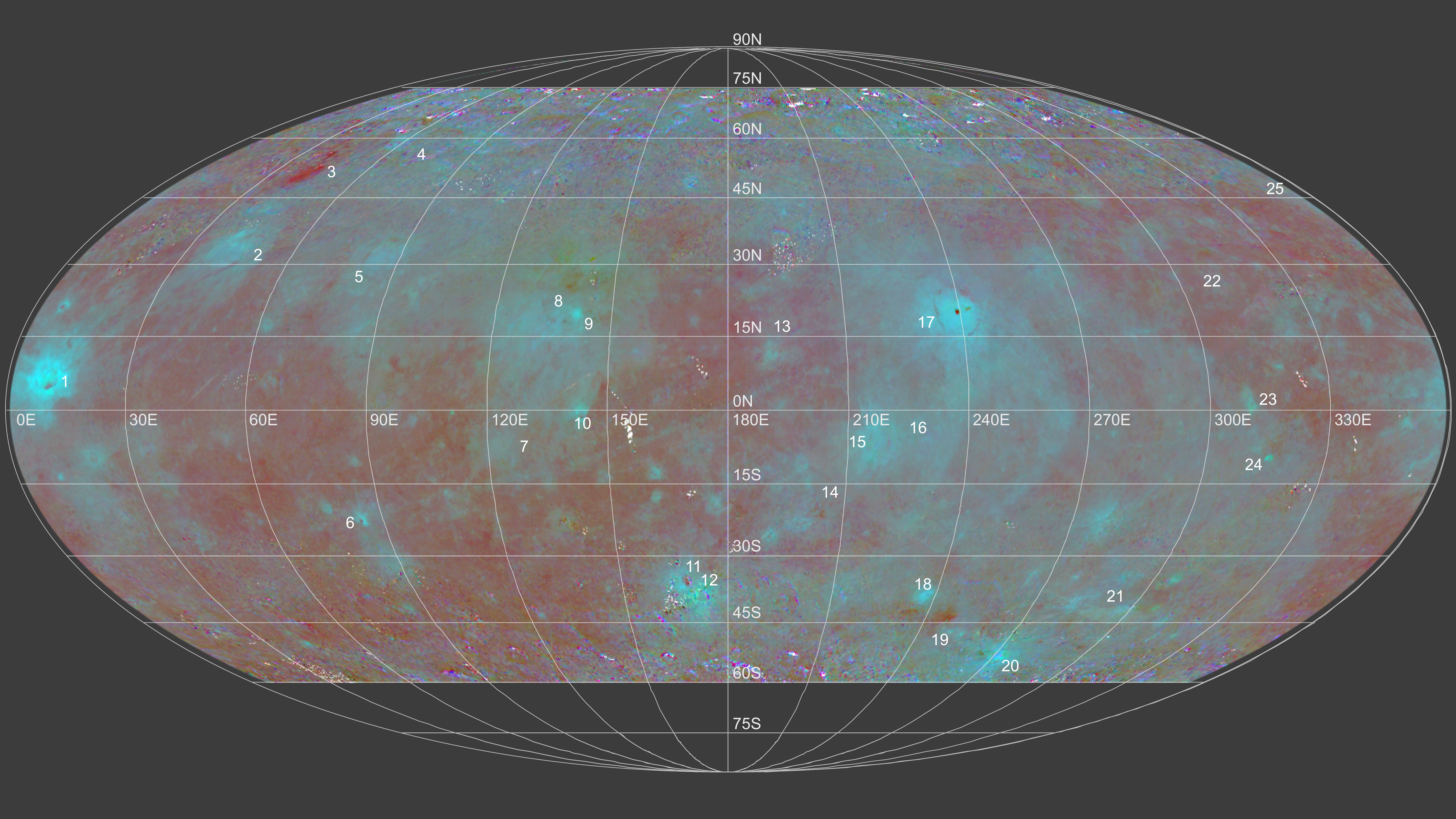}
    \caption{\label{Fig_RGB_ratio} VIR color composite map made using $950/749$, $550/749$, and $438/749$ ratios for the red, green, and blue channels, respectively. Numbers refer to the features of Table \ref{Table2}, as discussed in the text. White areas correspond to missing data.}
\end{figure*}

Figures \ref{Fig_RGB} and \ref{Fig_RGB_ratio} are two RGB-color composite maps. The same cubes as for Fig. \ref{Fig_IoF_550} have been filtered out. Some artifacts are still visible in Fig. \ref{Fig_RGB} (the same as in Fig. \ref{Fig_IoF_550}) but disappear in Fig. \ref{Fig_RGB_ratio} due to the use of ratios.\par
The first color map (Fig. \ref{Fig_RGB}) allows the variation of color and reflectance across the surface to be highlighted, providing additional information that is complementary to the $I/F_{550nm}$ map. Most of the Ceres surface appears to be gray-beige. Several blue features stand out across the surface. The most notable of these are the craters Haulani, Oxo, Occator (we note that the faculae inside the crater are overexposed and appear white instead of red), Juling, and Kupalo. Some features such as Ikapati, Centeotl, Braciaca, Ahuna Mons, and Tawals are also blue, but they are less evident because of their size or because their reflectance is lower. Vendimia Planitia is sufficiently visible in Fig. \ref{Fig_RGB}, which is due to its reflectance rather than its color. Dantu is the most striking feature in Vendimia Planitia and Fig. \ref{Fig_RGB} helps with visualizing a north-south dichotomy that is also visible in terms of color, with the southern part being bluer than the northern part (which encompasses the major part of the crater floor). The small blue dot visible in the southeast of the Dantu crater floor is the Centeotl crater. Very few red features stand out among the main gray-beige color and blue areas. The most evident is the reddish area observed towards the southwest and the northwest of crater Ernutet. We can potentially recognize the same red color in the Juling crater floor as well as on the northeast side of the central peak of Urvara crater. Figure \ref{Fig_RGB_ratio}  confirms and illustrates these observations more effectively.\par
The second color map (Fig. \ref{Fig_RGB_ratio}), based on the same RGB combination but normalized by the reflectance at \unit{749}{\nano\meter}, highlights only the color variations. It is then easier to identify differences in the surface properties, whether they are due to the composition or physical in origin. In Fig. \ref{Fig_RGB_ratio}, areas which were gray-beige in Fig. \ref{Fig_RGB} appear red and light blue. In this case, such colors highlight different units, light blue being clearly correlated to the crater ejecta and red denoting a background-like unit. The RGB ratio then allows us to better exploit those differences in comparison to the classic RGB map (Fig. \ref{Fig_RGB}), where only the most intense blue areas of Fig. \ref{Fig_RGB_ratio} are visible. This is well illustrated by the thin ejecta ray beginning at Occator and crossing half of Ceres' surface in the southwestern direction, roughly until the small Braciaca crater. Other ejecta rays, namely, of Occator and Haulani, are also sufficiently visible in Fig. \ref{Fig_RGB_ratio}, whereas they are not in the classic RGB map (Fig. \ref{Fig_RGB}). The bluer features through the RGB ratio correspond to Haulani, Oxo, and Occator. Juling and Kupalo stand out too, but the lack of data in that area and the vicinity of the pole imply more artifacts. Several small craters are also particularly visible: Centeotl, Braciaca, Cacaguat, Tawals, Nunghui, and an unnamed crater ($7.7\degr$E-$20.5\degr$N) north of Haulani. The red material of the Ernutet area is very well visible in Fig. \ref{Fig_RGB_ratio}. Red colors of equal intensity are also visible on the central peak of Urvara. A similar red color (though less intense) is observed on the respective floors of Juling and Braciaca, as well as on the unnamed crater mentioned above ($7.7\degr$E-$20.5\degr$N). In Fig. \ref{Fig_RGB_ratio}, Ahuna Mons appears green to green-blue. A bit farther north, the Xevioso crater exhibits the same green-blue color, particularly on its south ejecta. Finally, another tone of green is also identifiable in the northern region of Dantu.
%
\subsection{Map of the $S_{405-465nm}$ slope}
\label{SubSec_Map slope 405-465nm}
\begin{figure*}
    \centering
    \includegraphics[width=17cm]{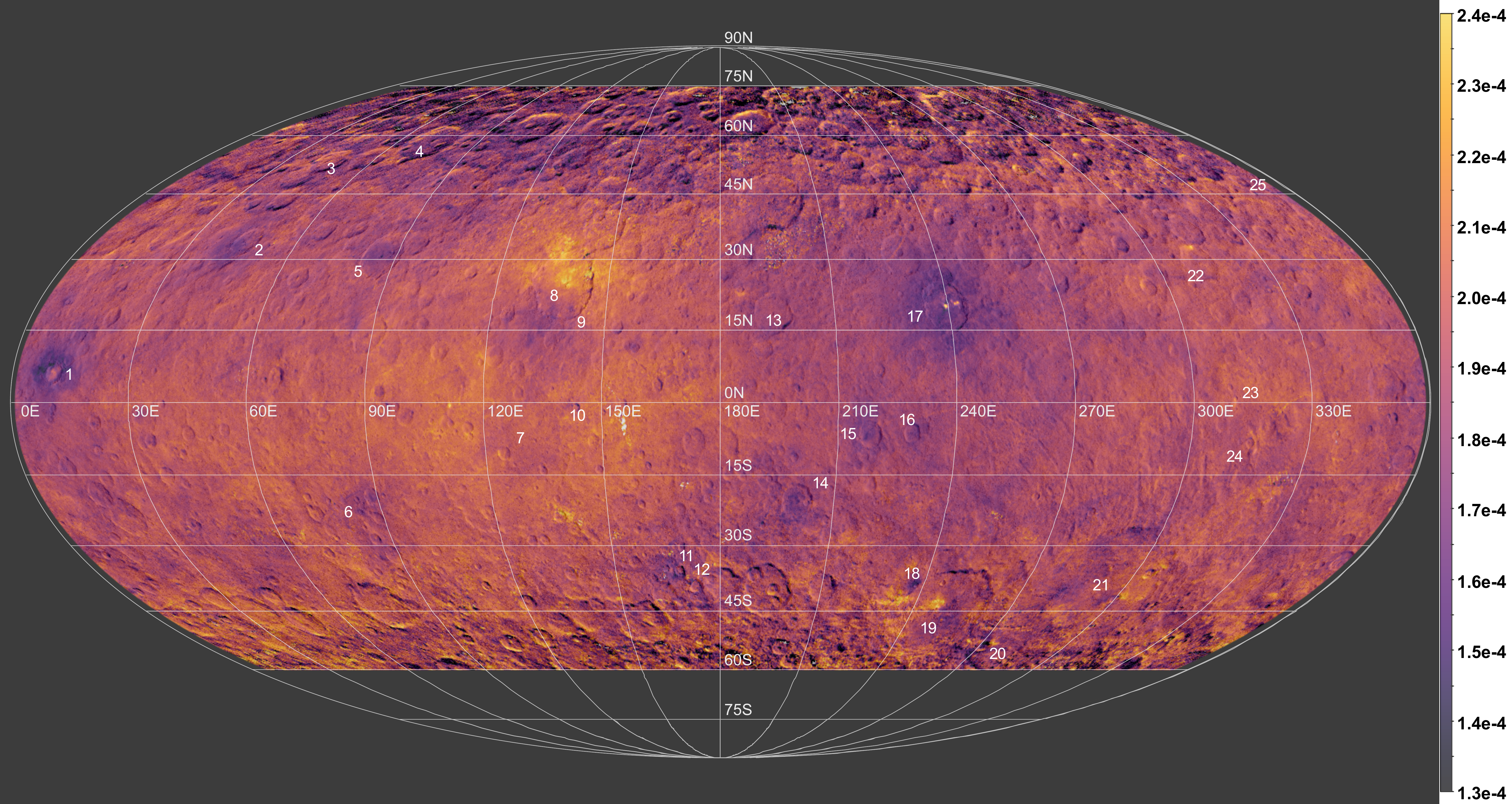}
    \caption{\label{Fig_SLOPE_405_465} Map of the VIR $S_{405-465nm}$ spectral slope superimposed over the FC LAMO map (see Sect. \ref{SubSec_Map projections}). Numbers refer to the features of Table \ref{Table2}, as discussed in the text. White areas correspond to missing data.}
\end{figure*}

The map of the slope between \unit{405}{\nano\meter} and \unit{465}{\nano\meter} is presented in Fig. \ref{Fig_SLOPE_405_465}. A total of 12 cubes (over 505) have been filtered out to avoid some major artifacts. We enhanced the contrast by choosing a color range in which 92$\%$ of the dataset falls (taking into account the 12 filtered-out cubes) to better highlight the variations of the slope observed on the surface. The values range from $1.3\times10^{-4}$ \reciprocal{\kilo\angstrom} to $2.4\times10^{-4}$ \reciprocal{\kilo\angstrom}, corresponding to 85$\%$ of the variation.\par
We observe a certain level of variability across the surface. The most evident feature at large scale, also visible in Fig. \ref{Fig_IoF_550}, is the difference between Vendimia Planitia and Hanami Planum (including the Occator crater), with the former characterized by a more positive slope on average. Within those two regions, we can distinguish different structures.\par
The Dantu crater, on Vendimia Planitia, is particularly visible through $S_{405-465nm}$ and is divided into two areas between the North and the South, as noted in Figs. \ref{Fig_IoF_550}, \ref{Fig_RGB}, and \ref{Fig_RGB_ratio} (see also Sect. \ref{subSec_Dantu}). The Haulani and Occator craters, which are known to be very complex (e.g., \cite{2018_Krohn, 2018_Tosi, 2019_Tosi, 2018_Schenk, 2019_Raponi_b, 2020_De_Sanctis_b}), are the two others Cerean features visible at large scale in Fig. \ref{Fig_SLOPE_405_465}. They are presented in detail in Sects. \ref{subSec_Occator} and \ref{subSec_Haulani}. Other features, such as craters Azacca, Lociyo, Ikapati, and Gaue, can be recognized. They all show a lower value of $S_{405-465nm}$, comprised between $1.69\times10^{-4}$ \reciprocal{\kilo\angstrom} and $1.77\times10^{-4}$ \reciprocal{\kilo\angstrom}, than the mean Ceres ($1.92\times10^{-4}$ \reciprocal{\kilo\angstrom}, see Fig. \ref{Fig_MED_SPEC}). We also note a bright spot on the northern rim of the Fejokoo crater, which is well visible with a slope around $2.33\times10^{-4}$ \reciprocal{\kilo\angstrom}. This area is also reported in studies of the bright spots by \cite{2019_Stein} and on the carbonate map by \cite{2018_Carrozzo}.\par
In the southern hemisphere, we distinguish a yellowish area corresponding to the unnamed crater centered at $138\degr$E--$24\degr$S. The central peak and the western rim of Urvara also exhibits a high positive slope (around $2.27\times10^{-4}$ \reciprocal{\kilo\angstrom}). Finally, the two yellowish spots inside Yalode correspond to the well-preserved Besua and Lono craters ($300\degr$E--$42.5\degr$S and $304\degr$E--$36.5\degr$S, see \cite{2018_Crown}), but they may be photometric artifacts.
%
\subsection{Map of the $S_{480-800nm}$ slope}
\label{SubSec_Map slope 480-800nm}
\begin{figure*}
    \centering
    \includegraphics[width=17cm]{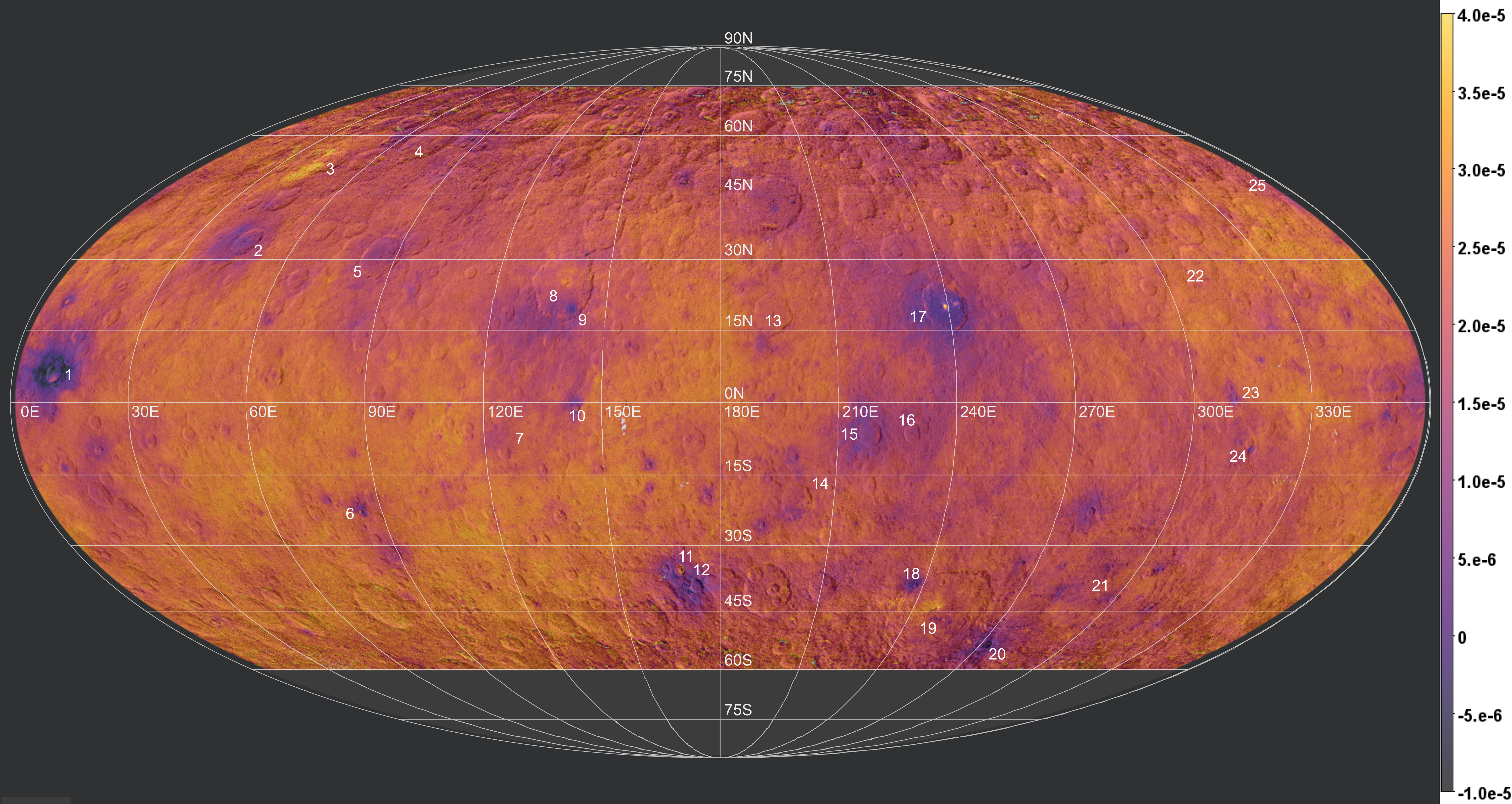}
    \caption{\label{Fig_SLOPE_480_800} Map of the VIR $S_{480-800nm}$ spectral slope superimposed over the FC LAMO map (see Sect. \ref{SubSec_Map projections}). Numbers refer to the features of Table \ref{Table2}, as discussed in the text. White areas correspond to missing data.}
\end{figure*}

The map of Fig. \ref{Fig_SLOPE_480_800} represents the slope between \unit{480}{\nano\meter} and \unit{800}{\nano\meter}, that is, the relative changes in the slope in the main range of the VIR VIS spectra. Twenty-four cubes providing major artifacts have been filtered out. The color scale, combined with the filtering of the cubes, represents $93\%$ of the initial dataset. Observations out of scale are mainly located at extreme latitudes, where the photometric correction is less accurate and the signal is low, and few of them are within the Haulani crater or spread across the surface. They generally do not affect the quality of the map due to the good redundancy (see Sect. \ref{SubSec_Data correction and processing} and Supplementary Fig. \ref{Fig_Appendix_DENSITY_MAP}).\par
The median of the global distribution is $2.47\times10^{-5}$ \reciprocal{\kilo\angstrom} and this corresponds to a slightly positive slope. However, the map seems to be well divided between recognizable structures (violet), which exhibit slopes close to zero or negative ($-1.0\times10^{-5}$ \reciprocal{\kilo\angstrom} < $S_{480-800nm}$ < $1.0\times10^{-5}$ \reciprocal{\kilo\angstrom}), and an orange background unit above $3.0\times10^{-5}$ \reciprocal{\kilo\angstrom} in slope. The violet units are generally impact craters or peculiar geologic formations which are enhanced with this spectral criterion. A couple of areas also show larger spectral slopes with respect to the average.\par
The most recognizable structures on the map are Oxo (not visible on the map presented here), Haulani, Centeotl (on the Dantu floor), Occator, Juling, Kupalo, Tawals, and Nunghui craters, which all show negative slopes. In contrast with the previous maps, Vendimia Planitia is not evident in this map. Similarly, the Dantu dichotomy is not discernible, although it is in Figs. \ref{Fig_IoF_550}, \ref{Fig_RGB}, and \ref{Fig_RGB_ratio}, where only the southern ejecta are visible. As observed in the RGB ratio map (Fig. \ref{Fig_RGB_ratio}), the pattern of ejecta and rays from Occator and Haulani are highlighted on that map, showing violet tones. In particular, the ejecta rays from Occator show mainly curved trajectories, while the ones from Haulani are straight and less extended. The exception is the thin ray coming out of Occator and crossing half of the Ceres surface in the southwestern direction.\par
The bright yellow area in the north of Ikapati corresponds to the Ernutet crater and the nearby southwest terrains. Just as for $S_{405-465nm}$, the central peak of Urvara is properly visible at a large scale and exhibits a slope, $S_{480-800nm}$ , as high as $3.25\times10^{-5}$ \reciprocal{\kilo\angstrom}.
%
\subsection{Map of the $S_{800-950nm}$ slope}
\label{SubSec_Map slope 800-950nm}
\begin{figure*}
    \centering
    \includegraphics[width=17cm]{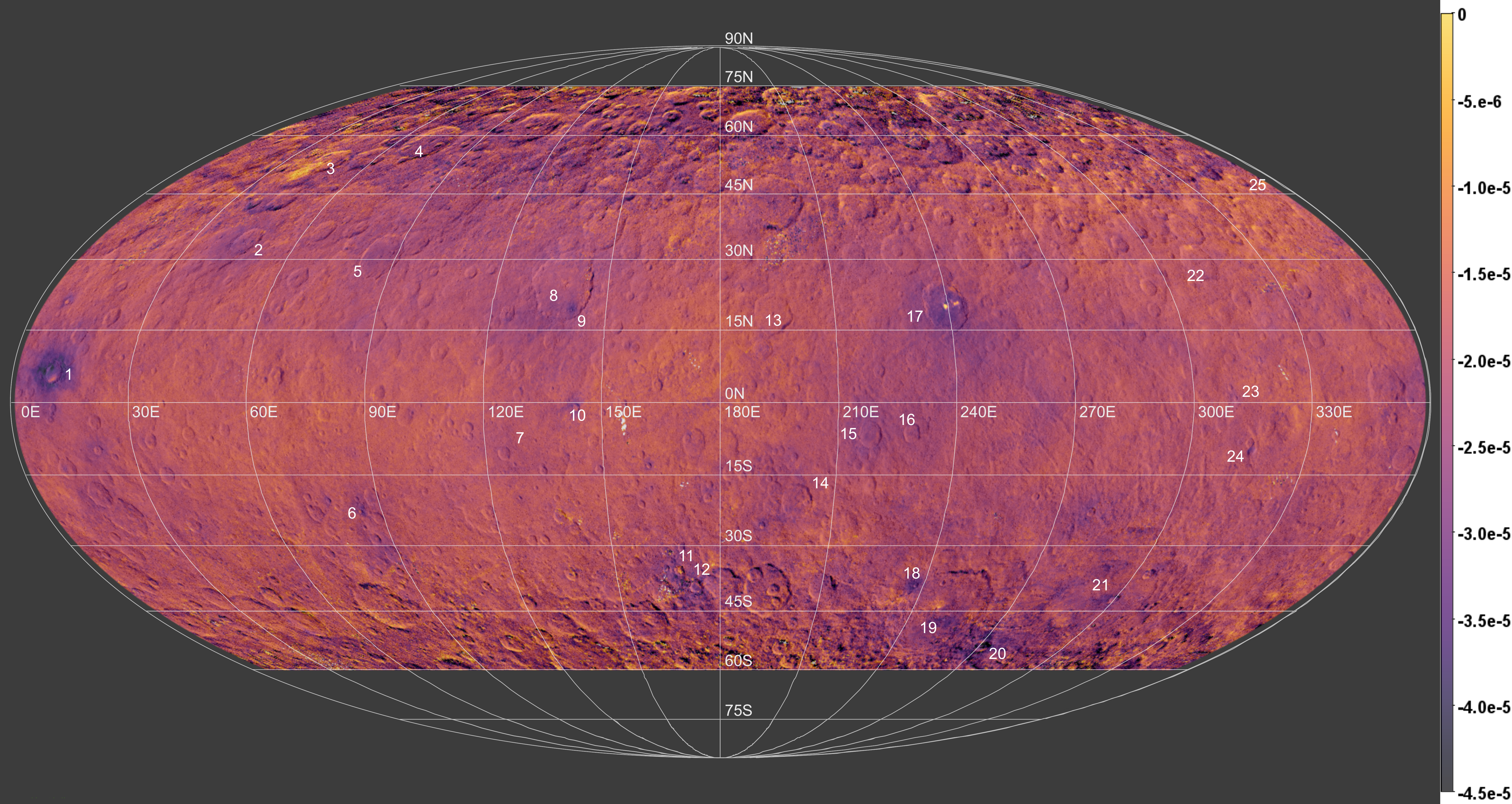}
    \caption{\label{Fig_SLOPE_800_950} Map of the VIR $S_{800-950nm}$ spectral slope superimposed over the FC LAMO map (see Sect. \ref{SubSec_Map projections}). Numbers refer to the features of Table \ref{Table2}, as discussed in the text. White areas correspond to missing data.}
\end{figure*}

The $S_{800-950nm}$ spectral indicator characterizes the near-IR part of the VIR visible channel. In that range, the spectral slope is negative and presents variations on Ceres' surface, illustrated by the map in Fig. \ref{Fig_SLOPE_800_950}. Forty-five cubes have been dismissed because they add artifacts to the map. With this filtering and a color scale between $-4.5\times10^{-5}$ \reciprocal{\kilo\angstrom} and $0$ \reciprocal{\kilo\angstrom}, it represents 87$\%$ of the global dataset.\par
The $S_{800-950nm}$ map brings out fewer details and less contrast than the $S_{480-800nm}$ map (Fig. \ref{Fig_SLOPE_480_800}), both at a global and local scale. Despite this, several differences are visible in the maps, especially related to large and complex craters.\par
The Haulani and Occator craters are the most visible craters on the surface and have the same properties as in the $S_{480-800nm}$ map, showing the most negative slopes. While it can be seen in the maps in Figs. \ref{Fig_SLOPE_405_465} and \ref{Fig_SLOPE_480_800}, here, the Dantu crater is nearly invisible. However, the Centeotl crater, which is located in the southeast floor of Dantu, shows strong negative slopes (close to $-4.02\times10^{-5}$ \reciprocal{\kilo\angstrom} for the crater floor and $-2.45\times10^{-5}$ \reciprocal{\kilo\angstrom} for the closest ejecta) and is properly visible even at a global scale. Ahuna Mons is also still visible through this spectral indicator, despite its relative small size (\unit{17}{\kilo\meter}).
As in Fig. \ref{Fig_SLOPE_480_800}, the Ernutet region is easily recognizable in Fig. \ref{Fig_SLOPE_800_950}, with a slope around $-6.47\times10^{-6}$ \reciprocal{\kilo\angstrom}. It is thus one of the highest areas on the surface.
%
\section{Discussion on areas of interest}
\label{Sec_Area_of_interest}
In this section, we describe and discuss the characteristics of features that stand out on the maps of the slopes. The main ones are Ahuna Mons, Dantu, Ernutet, Haulani, Juling, Kupalo, Occator, and Oxo (from Sect. \ref{subSec_Ahuna} to Sect. \ref{subSec_Oxo}). A figure is attached to each of the following section and is made up of: 1) a FC clear filter image extracted from the global LAMO mosaic \citep{2016_Roatsch_c} in a spherical projection; 2) three images corresponding to  $S_{405-465nm}$, $S_{480-800nm}$ , and $S_{800-950nm}$ in a spherical projection. The FC image is superimposed and color scales are the same as in Sect. \ref{Sec_Maps of the spectral parameters}; 3) different spectra extracted from the VIR data and numbered, as indicated on the spectral slope close-up. The spectra are a median of the observations available from our dataset on the selected region of interest to increase the signal-to-noise ratio. The spectral slope values provided in the text are calculated from these spectra.
\subsection{Ahuna Mons}
\label{subSec_Ahuna}
\begin{figure*}
    \centering
    \includegraphics[width=17cm]{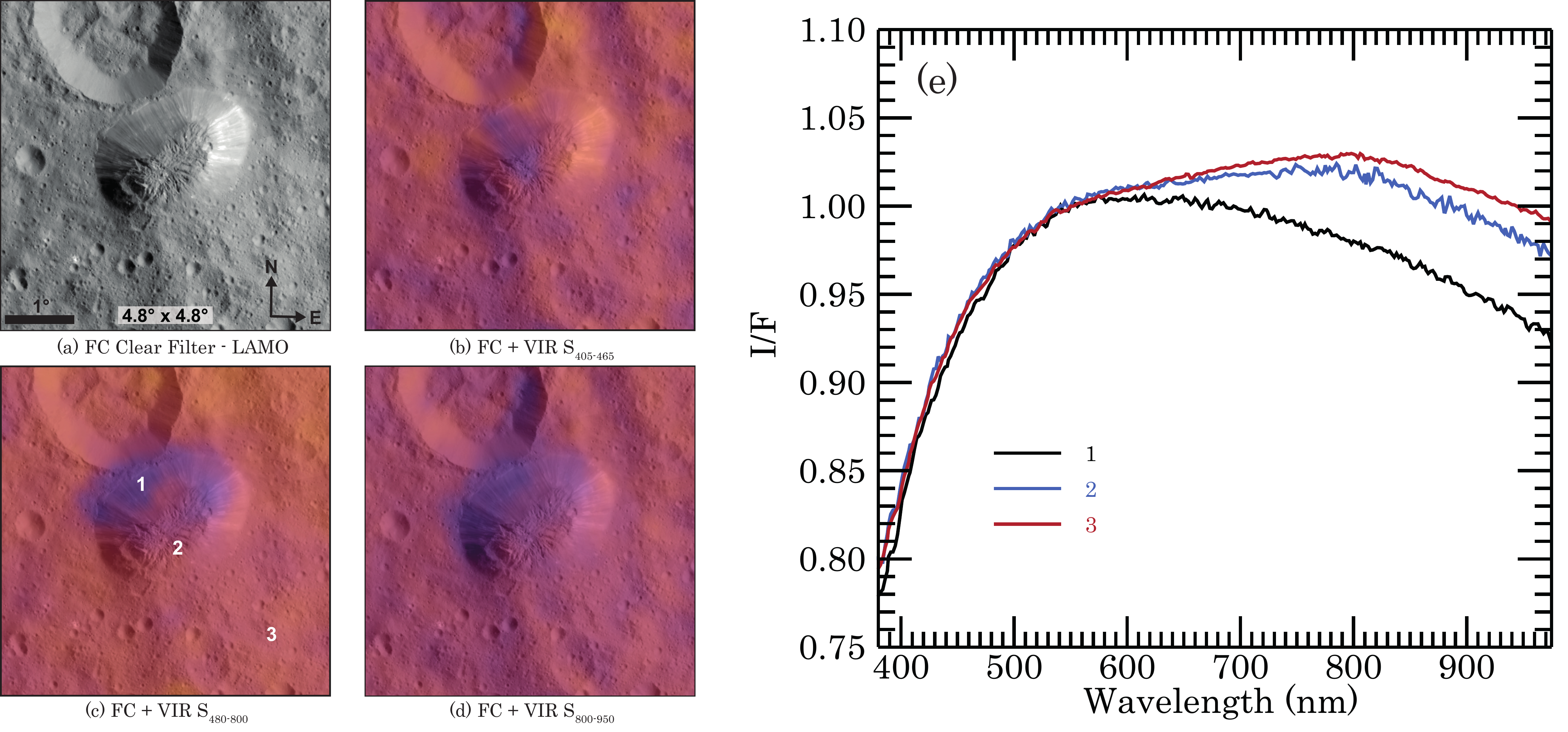}
    \caption{\label{Fig_Ahuna_Mons_Zoom} Close-up of Ahuna Mons and associated spectra. Ahuna Mons is \unit{17}{\kilo\meter}-wide. All the images use a spherical projection with a field of $4.8\degr$ by $4.8\degr$ and have the same orientation. (a) Framing Camera clear filter image from the LAMO mission phase; (b) VIR $S_{405-465nm}$; (c) VIR $S_{480-800nm}$ with indication of the regions of interest where spectra of panel (e) have been extracted; (d) VIR $S_{800-950nm}$; (e) spectra from the regions of interest of panel (c) normalized at \unit{550}{\nano\meter}.}
\end{figure*}

Ahuna Mons is a \unit{17}{\kilo\meter}-wide, \unit{4}{\kilo\meter}-high mountain of cryovolcanic origin \citep{2016_Ruesch}. Its recent emplacement at geological timescale (not older than $210\pm30$ million years) is likely due to the ascent of a slurry cryomagma \citep{2016_Ruesch, 2019_Ruesch}. The flanks of Ahuna Mons, where linear features are visible, are $30\degr$ to $40\degr$ steep, and the top is made of fractured and hummocky terrains \citep{2016_Ruesch}. In Fig. \ref{Fig_Ahuna_Mons_Zoom}, $S_{405-465nm}$ does not show significant spatial variability (see also the regional context on Fig. \ref{Fig_SLOPE_405_465}). Slopes $S_{480-800nm}$ and $S_{800-950nm}$ have sharper variations than $S_{405-465nm}$. In particular, the northwestern and the north-northeastern flanks are characterized by smaller values of $S_{480-800nm}$ slope and $S_{800-950nm}$ (spectrum n$\degr1$; $S_{480-800nm}\simeq7.49\times10^{-6}$ \reciprocal{\kilo\angstrom} and $S_{800-950nm}\simeq-2.97\times10^{-5}$ \reciprocal{\kilo\angstrom}) than the surrounding terrains (spectrum n$\degr3$; $S_{480-800nm}\simeq2.18\times10^{-5}$ \reciprocal{\kilo\angstrom} and $S_{800-950nm}\simeq-2.04\times10^{-5}$ \reciprocal{\kilo\angstrom}). The spectral behavior of the top and of the southeast flank of Ahuna Mons is closer to the surrounding terrains with $S_{480-800nm}\simeq1.76\times10^{-5}$ \reciprocal{\kilo\angstrom} and $S_{800-950nm}\simeq-2.49\times10^{-5}$ \reciprocal{\kilo\angstrom}.\par
The infrared observations of the VIR have allowed for the identification of sodium carbonates on the flanks of Ahuna Mons \citep{2017_Zambon, 2018_Carrozzo, 2019_Zambon}, especially from the west to the northeast flanks (clockwise). The spectral slope calculated with the FC data in \cite{2019_Zambon} are qualitatively in agreement with our results for the $S_{480-800nm}$. Thanks to the high resolution of the FC, \cite{2016_Ruesch} and \cite{2018_Platz} report that brighter areas and linear features are present on the same flanks. Those linear features imply a mass-wasting, with material falling down from the top of the mons, while this is not the case on the southeastern flanks, which are less steep \citep{2018_Platz}. Those observations are consistent with the spectral slope variations derived by VIR data. In particular, the reduction of the spectral slope in the northwestern-north and northeastern flanks are compatible with the presence of fresher material or carbonates exposed in the mass-wasting.
%
\subsection{Dantu}
\label{subSec_Dantu}
Dantu is a large complex crater, \unit{126}{\kilo\meter} in diameter, which is located in the northern part of Vendimia Planitia. The datation of the ejecta material of Dantu indicates a formation epoch between $25$ and $150$ million years \citep{2018_Williams_b}. The complex geomorphological context of Dantu has been studied in detail through the high-resolution FC images by, for example, \cite{2016_Kneissl, 2018_Williams_b, 2018_Stephan, 2019_Stephan}, while the mineralogical context is detailed in \cite{2019_Stephan}.\par
\begin{figure*}
    \centering
    \includegraphics[width=17cm]{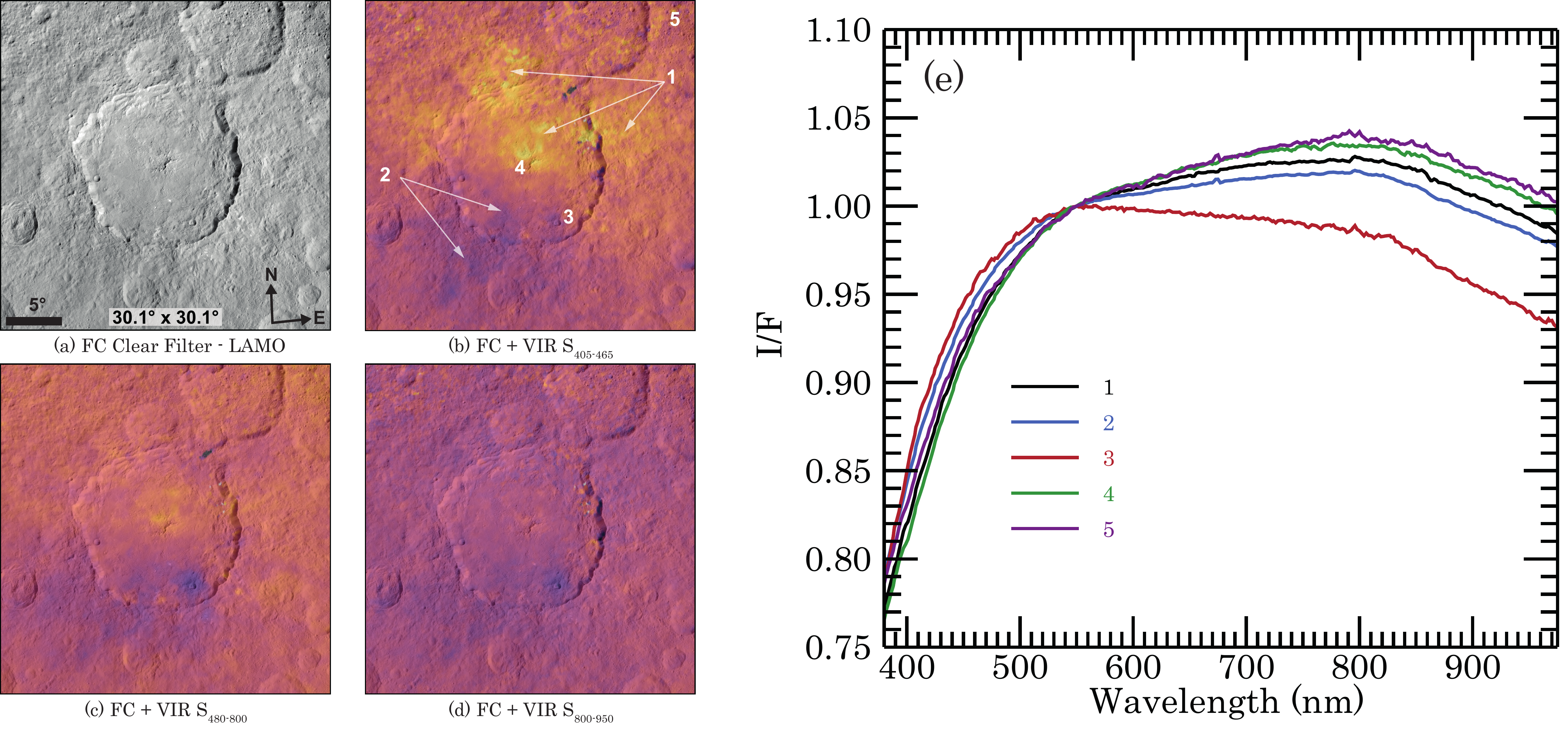}
    \caption{\label{Fig_Dantu_Zoom} Same as Fig. \ref{Fig_Ahuna_Mons_Zoom} for Dantu. Dantu has a diameter of \unit{126}{\kilo\meter}. All the images use a spherical projection with a field of $30.1\degr$ by $30.1\degr$ and have the same orientation.}
\end{figure*}

As we show in Fig. \ref{Fig_Dantu_Zoom}, the Dantu region presents unique characteristics on the surface of Ceres. Throughout $S_{405-465nm}$, which represents the majority of the slope variations in this region, the Dantu crater is divided into two distinct parts. The northern floor and ejecta show several different properties compared to the southern ejecta, which behave like other crater ejecta. In particular, the northern region of the crater presents one of the steepest slopes of the surface (see Fig. \ref{Fig_SLOPE_405_465} and area and spectrum n$\degr1$ of Fig. \ref{Fig_Dantu_Zoom}; $S_{405-465nm}\simeq2.18\times10^{-4}$ \reciprocal{\kilo\angstrom}), which is also greater than their southern counterpart (area and spectrum n$\degr2$; $S_{405-465nm}\simeq1.89\times10^{-4}$ \reciprocal{\kilo\angstrom}). The reflectance ($I/F_{550nm}$) map (Figs. \ref{Fig_IoF_550}) also shows this north-south dichotomy, with the south-southwest region of the crater floor and ejecta being darker ($I/F_{550nm}\simeq0.034$) than the northern part ($I/F_{550nm}\simeq0.037$). In addition, a beige and a green color characterize the north of Dantu on the RGB composites, in Figs. \ref{Fig_RGB} and \ref{Fig_RGB_ratio}, respectively, while other ejecta are generally blue (see also \cite{2017_Schroder}).\par
The central peak of Dantu is complex, presenting a partially collapsed morphology on its west side and fractured terrains on the northern part \citep{2018_Stephan}. It presents the steepest $S_{405-465nm}$ slope of Ceres (area and spectrum n$\degr4$; $S_{405-465nm}\simeq2.29\times10^{-4}$ \reciprocal{\kilo\angstrom}), and it is also distinguishable on the $S_{480-800nm}$ map (area and spectrum n$\degr4$; $S_{480-800nm}\simeq2.73\times10^{-5}$ \reciprocal{\kilo\angstrom}).\par
Throughout the $S_{480-800nm}$ and $S_{800-950nm}$ slopes, the Dantu crater itself does not present characteristics that are seen as well as through the $I/F_{550nm}$, the $S_{405-465nm}$, and the RGB maps. The main remarkable feature here is the young Centeotl crater, located in the southeast of Dantu's floor, which shows an almost flat $S_{480-800nm}$ slope (area and spectrum n$\degr3$; $S_{480-800nm}\simeq4.99\times10^{-5}$ \reciprocal{\kilo\angstrom}); the part of the spectrum beyond \unit{500}{\nano\meter} is, nevertheless, negative in slope.\par
Based on the mineralogical analysis from \cite{2016_Ammannito} and \cite{2018_Stephan}, we observe a strong correlation between the band depths at \unit{2.7}{\micro\meter} and \unit{3.1}{\micro\meter,} corresponding to the structural \ce{OH} and \ce{NH4} absorptions in phyllosilicates, respectively -- and our $S_{405-465nm}$ spectral slope and, to a lesser extent, with the $S_{480-800nm}$ slope. The variation of dark phase abundance or grain size is not likely considered to explain the global variations of the \unit{3.1}{\micro\meter} band on Ceres \citep{2016_Ammannito}. However, this cannot be excluded in the case of Dantu given the uniqueness of the VIR VIS observations.\par 
The global map of \cite{2018_Carrozzo} revealed localized areas richer in carbonates, where at least one of them is associated with a bright spots \citep{2019_Palomba_b}. However, the distribution of carbonates is very localized and the north-south dichotomy observed in our spectral slope is not visible in the carbonates distribution.\par
Thus, the variations observed in that region with the VIR VIS channel strongly support a change in composition or in the physical properties of the surface, which is not observed elsewhere on the Ceres surface. However, based on: 1) the absence of link between our observations and the carbonates distributions; 2) the absence of identification of the mineral species (one or more) responsible for the UV-VIS absorption in the Ceres spectrum; 3) the observations of \cite{2016_Ammannito}, who excluded a different nature in the phyllosilicates across the surface due to the absence of an observed shift in the phyllosilicate band, thus, the origin of the Dantu characteristics remains an open question.
%
\subsection{Ernutet}
\label{subSec_Ernutet}
\begin{figure*}
    \centering
    \includegraphics[width=17cm]{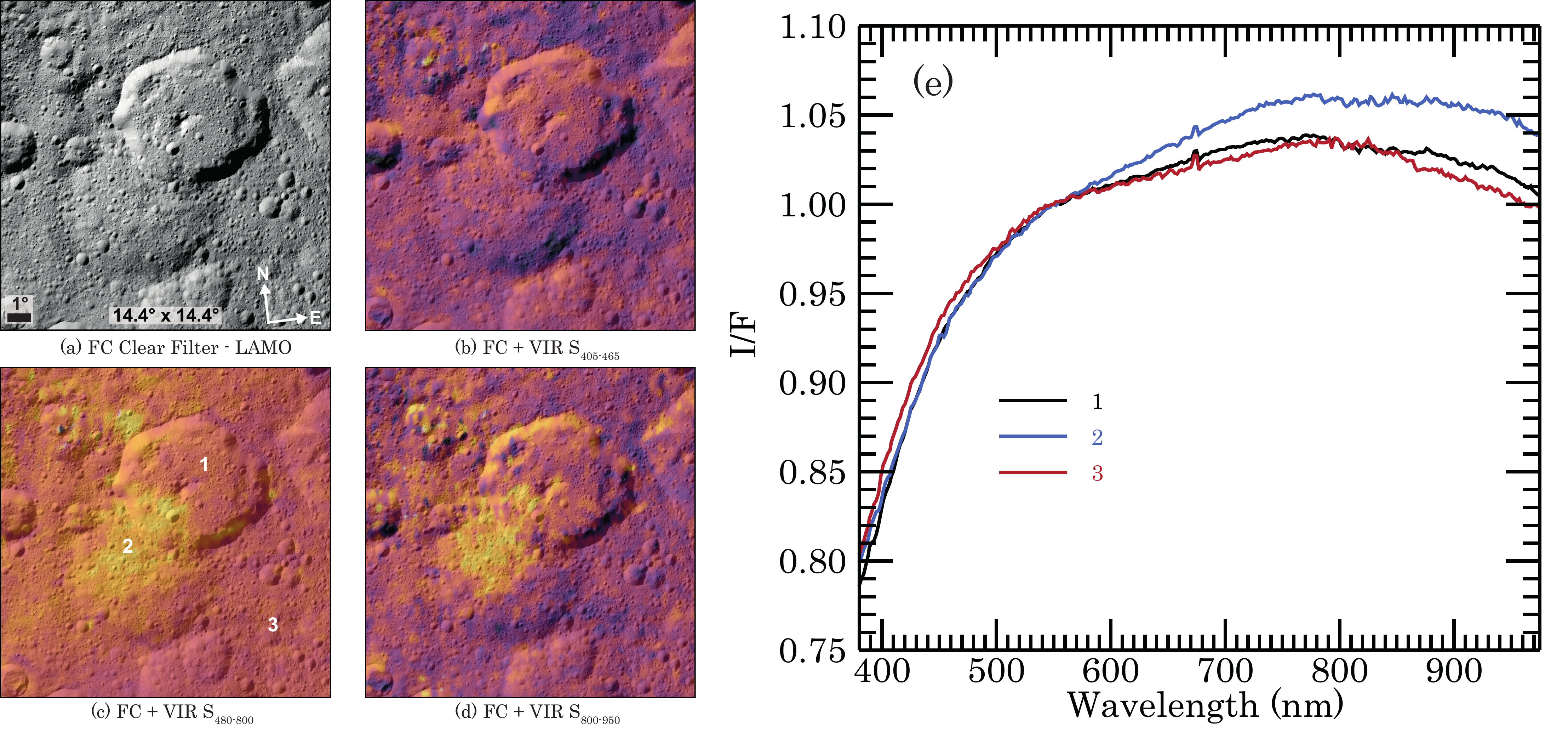}
    \caption{\label{Fig_Ernutet_Zoom} Same as Fig. \ref{Fig_Ahuna_Mons_Zoom} for Ernutet. Ernutet has a diameter of \unit{52}{\kilo\meter}. All the images use a spherical projection with a field of $14.4\degr$ by $14.4\degr$ and have the same orientation.}
\end{figure*}

Ernutet is a \unit{52}{\kilo\meter}-diameter crater located in the high latitudes of the northern hemisphere of Ceres that straddles an older crater of about the same size. Ernutet is the place on Ceres with the most evident spectral signature of organic-rich materials, as detected by the VIR spectrometer, thanks to a distinctive feature between \unit{3.3}{\micro\meter} and \unit{3.6}{\micro\meter} \citep{2017_De_Sanctis, 2018_De_Sanctis_b}. The observations from VIR in the visible range reveal a larger spectral slope on the same areas corresponding to the presence of organic-rich material, as illustrated by spectrum n$\degr2$ in Fig. \ref{Fig_Ernutet_Zoom}, in agreement with Framing Camera observations \citep{2017_Pieters_b}. Those areas correspond to an extended part in the southwest of Ernutet -- both on the crater floor and outside -- as well as on a smaller spot in the northwest \citep{2017_De_Sanctis, 2019_Raponi_a}.\par
Ernutet's organic-rich material areas clearly stand out in the maps of the $S_{480-800nm}$ and $S_{800-950nm}$ slopes, while it is not visible through the $S_{405-465nm}$ slope. On Ceres' surface, the area n$\degr2$ (Fig. \ref{Fig_Ernutet_Zoom}) has, with the Occator faculae, the highest values for $S_{480-800nm}$ ($\sim3.39\times10^{-5}$ \reciprocal{\kilo\angstrom}) and $S_{800-950nm}$ ($\sim-6.47\times10^{-6}$ \reciprocal{\kilo\angstrom}). For comparison, the Ernutet crater floor (area and spectrum n$\degr1$) and the control area n$\degr3$ in Fig. \ref{Fig_Ernutet_Zoom} have $S_{480-800nm}$ slope values of $\sim2.63\times10^{-5}$ \reciprocal{\kilo\angstrom} and $\sim2.32\times10^{-5}$ \reciprocal{\kilo\angstrom} , respectively, and $S_{800-950nm}$ value of $\sim-1.28\times10^{-5}$ \reciprocal{\kilo\angstrom} and $\sim-1.89\times10^{-5}$ \reciprocal{\kilo\angstrom} , respectively.\par
Previous VIR observations in the infrared indicates the presence of carbonates in the Ernutet areas \citep{2018_Carrozzo, 2019_Raponi_a}. Those carbonates are spread in the southwest area (corresponding to area n$\degr2$ in Fig. \ref{Fig_Ernutet_Zoom}), but they are not present in the small patch in the northwest area, as reported by \cite{2019_Raponi_a}. We notice that this is compatible with distribution of the $S_{800-950nm}$ slope. 
%
\subsection{Haulani}
\label{subSec_Haulani}
The \unit{34}{\kilo\meter}-diameter Haulani crater is one of the youngest craters of Ceres, with an age of formation no older than $\sim6$ million years, regardless of the model being used for its age estimation \citep{2016_Schmedemann, 2018_Krohn}. The geological and mineralogical complexity of Haulani \citep{2018_Krohn, 2018_Tosi, 2019_Tosi} has made it one of the most interesting features on the surface of the dwarf planet. The spectral criteria we adopt in the visible VIR also highlight this complexity.\par
First, the Haulani central peak and a part of the crater floor are one of the brightest areas of Ceres -- after the Occator faculae, and the Oxo and Kupalo craters -- with $I/F_{550nm}$ reaching 0.050 on the central peak. Furthermore, we note that the $I/F_{550nm}$ of the crater floor is relatively well correlated with the $I/F_{1.2\micro\meter}$ and the $I/F_{1.9\micro\meter}$ while on the ejecta, compared to the Ceres average, the $I/F_{550nm}$ is brighter and both the $I/F_{1.2\micro\meter}$ and the $I/F_{1.9\micro\meter}$ are darker \citep{2017_Ciarniello, 2018_Tosi, 2019_Tosi}. If not the brightest, Haulani is the bluest feature of the surface, as illustrated by Figs. \ref{Fig_RGB} and \ref{Fig_RGB_ratio}, and already noted by \cite{2016_Nathues, 2017_Stephan, 2017_Schroder, 2018_Tosi} and \cite{2019_Tosi}. This argues for the very young age of its formation \citep{2017_Stephan}.\par
\begin{figure*}
    \centering
    \includegraphics[width=17cm]{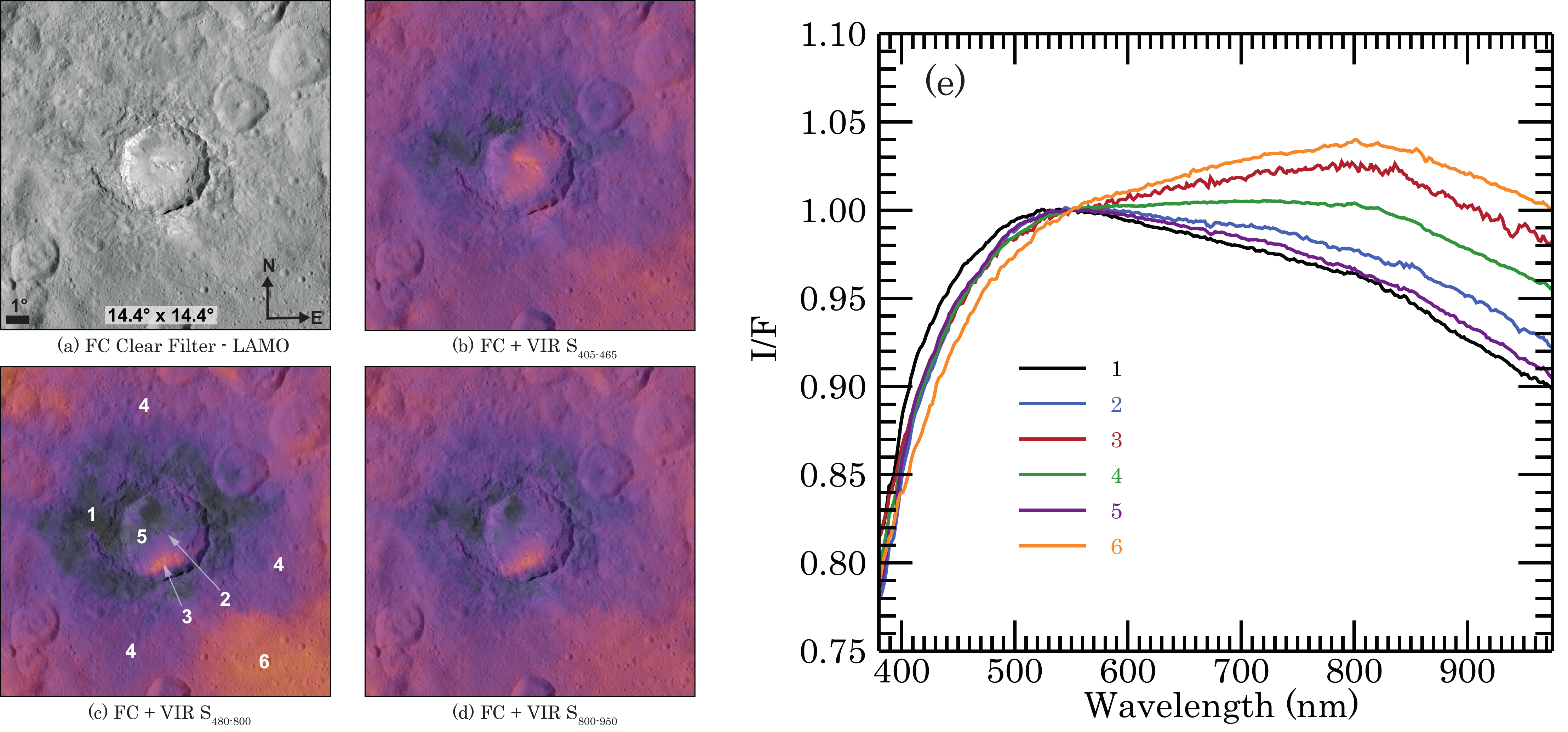}
    \caption{\label{Fig_Haulani_Zoom} Same as Fig. \ref{Fig_Ahuna_Mons_Zoom} for Haulani. Haulani has a diameter of \unit{34}{\kilo\meter}. All the images use a spherical projection with a field of $14.4\degr$ by $14.4\degr$ and have the same orientation.}
\end{figure*}

The three spectral slopes mapped in Fig. \ref{Fig_Haulani_Zoom} present different characteristics, the $S_{405-465nm}$ showing a different spatial distribution when compared with the $S_{480-800nm}$ and $S_{800-950nm}$ slopes, which are more similar to each other. The northern, western, and eastern rims of Haulani and the associated ejecta blanket (area and spectrum n$\degr1$ of Fig. \ref{Fig_Haulani_Zoom}) show the lowest value of the slope $S_{405-465nm}$ of the features presented in Sect. \ref{Sec_Area_of_interest}, with a value as low as $1.49\times10^{-4}$ \reciprocal{\kilo\angstrom}. The crater floor exhibits higher $S_{405-465nm}$ (around $1.90\times10^{-4}$ \reciprocal{\kilo\angstrom}), which are mainly correlated with the central peak, but nevertheless encompass a larger area that includes the south of the crater floor and a small patch on the northeastern talus material (see area and spectrum n$\degr2$).\par
The ejecta of Haulani, mostly westward-oriented, are adequately visible in the spectral slope maps (Figs. \ref{Fig_SLOPE_405_465}, \ref{Fig_SLOPE_480_800}, and \ref{Fig_SLOPE_800_950}), and, in $S_{480-800nm}$, show the largest contrast with respect to the surrounding terrains. Those latter are represented by the area and the spectrum n$\degr6$ in Fig. \ref{Fig_Haulani_Zoom}, have a slope $S_{480-800nm}$ around $1.65\times10^{-5}$ \reciprocal{\kilo\angstrom}. Conversely, the closest to the crater and bluest ejecta (n$\degr1$), as well as the one a bit farther (n$\degr4$), exhibit $S_{480-800nm}$ around $-6.42\times10^{-6}$ \reciprocal{\kilo\angstrom} and $9.58\times10^{-6}$ \reciprocal{\kilo\angstrom}, respectively.\par
Haulani crater's floor presents interesting features revealed by the spectral slopes $S_{480-800nm}$ and $S_{800-950nm}$. The north, illustrated by the spectrum n$\degr5$ on Fig. \ref{Fig_Haulani_Zoom}, shows a patch as blue as the bluest ejecta n$\degr1$. This patch corresponds to the "crater
floor material smooth dark" \textit{} (or cfsd unit), as mapped by \cite{2018_Krohn}. On the opposite side of the crater floor (area n$\degr3$), the $S_{480-800nm}$ and $S_{800-950nm}$ slopes are higher, reaching values of $1.65\times10^{-5}$ \reciprocal{\kilo\angstrom} and $-2.29\times10^{-5}$ \reciprocal{\kilo\angstrom}, respectively, which means relatively close to the surrounding area n$\degr6$. This peculiar zone (area n$\degr3$) corresponds to the "crater floor
material hummocky bright"  (or cfhb unit), as mapped by \cite{2018_Krohn} and which probably corresponds to material fallen down from the crater rims which are steeper on this side of the crater \citep{2018_Krohn}.\par
The Haulani crater, revealed through our spectral indices, can be compared with the composition maps that have already been published. We first note that no obvious correlation is observed between our spectral slopes and the carbonates distribution around Haulani \citep{2018_Carrozzo, 2019_Tosi}. However, a good qualitative correspondence can be drawn between the more negative $S_{480-800nm}$ slope and the lower band depth at \unit{2.7}{\micro\meter} and \unit{3.1}{\micro\meter} \citep{2016_Ammannito, 2019_Tosi}. A good qualitative correlation is also observed between the $S_{480-800nm}$ and $S_{800-950nm}$ slopes on area n$\degr3$ and the \unit{2.7}{\micro\meter} band depth, but less with the \unit{3.1}{\micro\meter} band depth \citep{2018_Tosi, 2019_Tosi}. Finally, the visible spectral slopes, $S_{480-800nm}$ and $S_{800-950nm}$ , are correlated with the infrared spectral slope, at least in the area n$\degr3$ and with the ejecta in general \citep{2018_Tosi, 2019_Tosi}.
%
\subsection{Juling and Kupalo}
\label{subSec_Juling and Kupalo}
\begin{figure*}
    \centering
    \includegraphics[width=17cm]{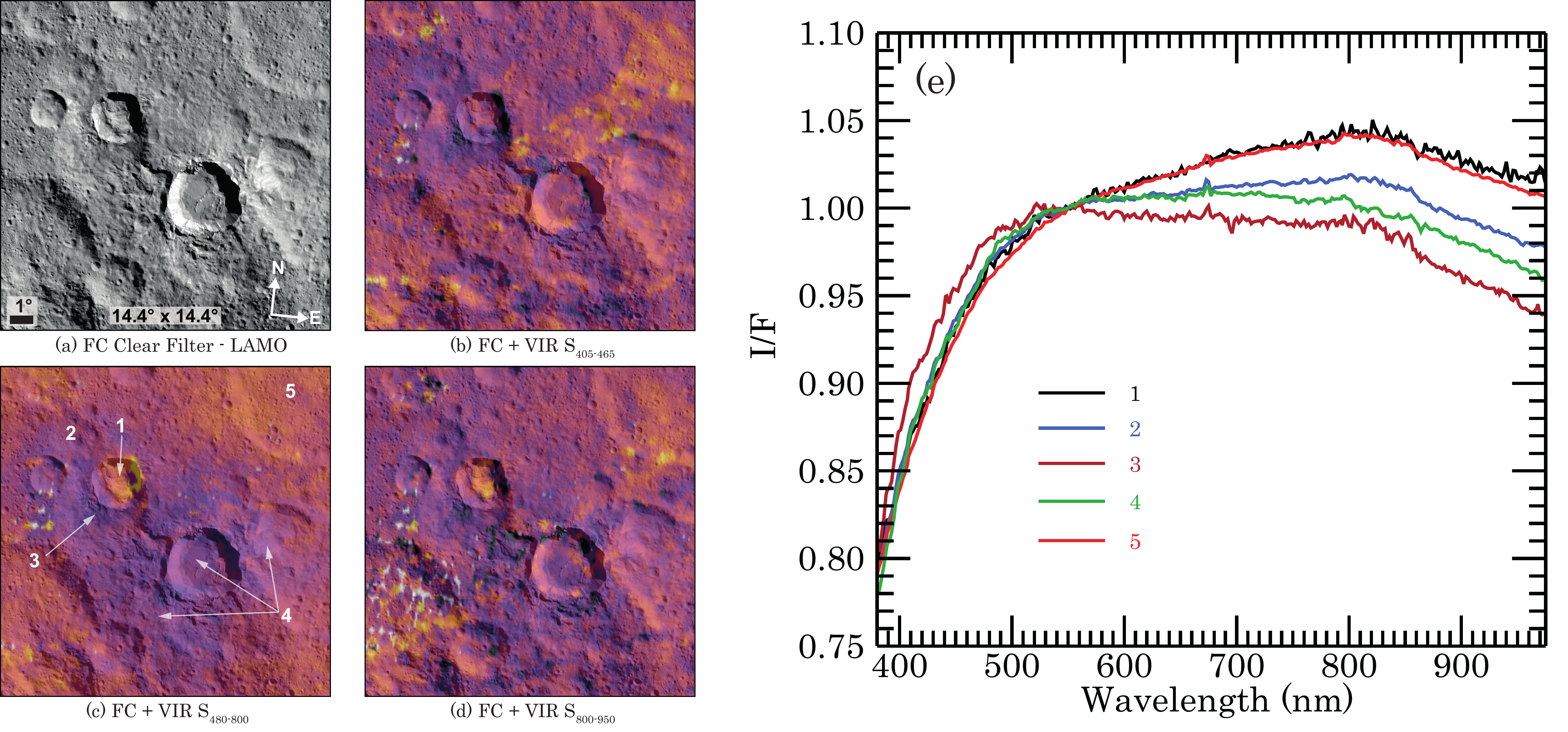}
    \caption{\label{Fig_Juling_Kupalo_Zoom} Same as Fig. \ref{Fig_Ahuna_Mons_Zoom} for Juling and Kupalo. Juling and Kupalo have diameters of \unit{20}{\kilo\meter} and \unit{26}{\kilo\meter}, respectively. All the images use a spherical projection with a field of $14.4\degr$ by $14.4\degr$ and have the same orientation.}
\end{figure*}

Juling and Kupalo are two neighbor craters of \unit{20}{\kilo\meter} and \unit{26}{\kilo\meter} in diameter; Kupalo, south of Juling, is younger. The analysis of the spectral diversity with our data is difficult due to the sparse coverage of this area. This is visible as white irregular dots on the spectral slope maps and noisy spectra in Fig. \ref{Fig_Juling_Kupalo_Zoom}.\par
While the context of Juling and Kupalo is very rich in detail (e.g., \cite{2017_Stephan,2019_De_Sanctis}), the visible spectral indices do not show a huge diversity. One remarkable area is the floor of Juling (area and spectrum n$\degr1$), which stands out in the $S_{480-800nm}$ and $S_{800-950nm}$ maps and has spectral slopes similar to the terrain outside the ejecta (area and spectrum n$\degr5$) with values around $2.48\times10^{-5}$ \reciprocal{\kilo\angstrom} and $2.65\times10^{-5}$ \reciprocal{\kilo\angstrom} for the $S_{480-800nm}$ and around $-1.68\times10^{-5}$ \reciprocal{\kilo\angstrom} and $-1.92\times10^{-5}$ \reciprocal{\kilo\angstrom} for the $S_{800-950nm}$, respectively. However, the reflectance of the two areas are quite different, with the Juling crater floor having a median $I/F_{550nm}$ around $0.039$, while the area n$\degr5$ close to 0.034, is similar to the mean of Ceres.\par
One recurrent feature observed in the three spectral slopes is a patch in the south west ejecta blanket of Juling, designated as area n$\degr3$. This area has the lowest values of each slope in Fig. \ref{Fig_Juling_Kupalo_Zoom} ($S_{405-465nm}\simeq1.56\times10^{-4}$ \reciprocal{\kilo\angstrom}, $S_{480-800nm}\simeq3.16\times10^{-6}$ \reciprocal{\kilo\angstrom} , and $S_{800-950nm}\simeq-3.06\times10^{-5}$ \reciprocal{\kilo\angstrom}).\par
Kupalo, in addition to being bright like Juling (median $I/F_{550nm}$ of $0.040$), appears blue in the maps of Figs. \ref{Fig_RGB} and \ref{Fig_RGB_ratio}, and as also pointed out by e.g., \cite{2016_Nathues, 2017_Stephan}. The low value of the slope $S_{480-800nm}$ ($1.25\times10^{-5}$ \reciprocal{\kilo\angstrom}) of the Kupalo crater floor and its ejecta (area n$\degr4$) corresponds to a mostly flat spectrum beyond \unit{500}{\nano\meter}. This is still bluer than the Juling ejecta (area and spectrum n$\degr2$) which exhibit $S_{480-800nm}$ around $1.53\times10^{-5}$ \reciprocal{\kilo\angstrom}.\par
Carbonates have been detected all around the Kupalo crater with the higher abundance in the south-west part \citep{2018_Carrozzo}. On the contrary, the band depths at \unit{2.7}{\micro\meter} and at \unit{3.1}{\micro\meter} are lower in those places, as shown by \cite{2016_Ammannito} and \cite{2019_De_Sanctis}. However, in both cases, we do not observe a peculiar spectral behavior at VIS wavelengths.
%
\subsection{Occator}
\label{subSec_Occator}
\begin{figure*}
    \centering
    \includegraphics[width=17cm]{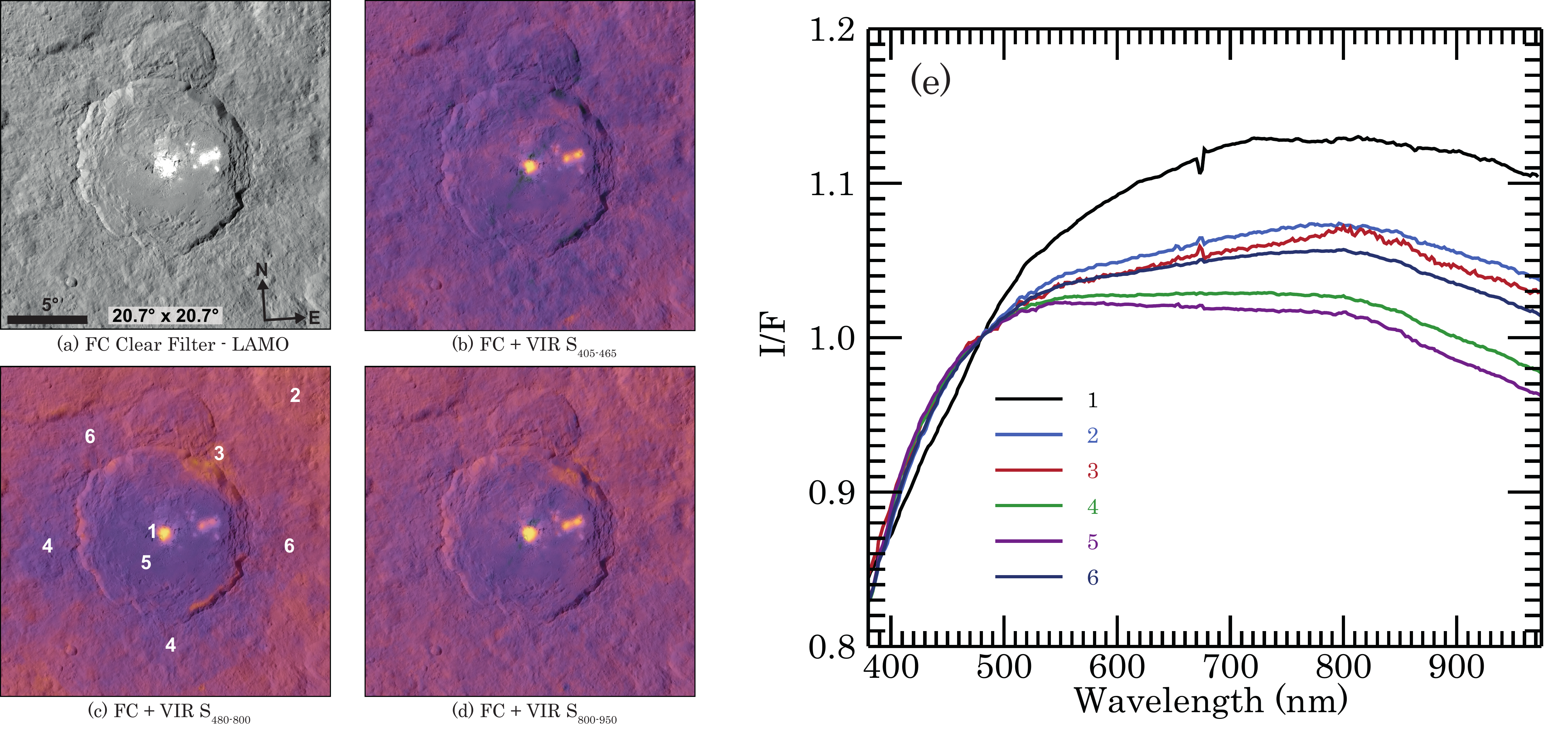}
    \caption{\label{Fig_Occator_zoom} Same as Fig. \ref{Fig_Ahuna_Mons_Zoom} for Occator. Occator has a diameter of \unit{92}{\kilo\meter}. All the images use a spherical projection with a field of $20.7\degr$ by $20.7\degr$ and have the same orientation.}
\end{figure*}

Occator (\unit{92}{\kilo\meter} in diameter) is one of the most interesting and intriguing craters of Ceres. Its peculiarity, as visible in Fig. \ref{Fig_Occator_zoom}, is attributed to two very bright spots, or faculae, of endogenous origin and which have been tentatively observed in some pre-Dawn Ceres observations \citep{2006_Li, 2008_Carry}. The origin and the formation of Occator have been abundantly discussed (see \cite{2019_Scullya} and references within). Concerning the central bright spot -- the complex Cerealia facula and dome -- the latest VIR IR observations suggests the presence of hydrated sodium chloride, which would have been able to lower the eutectic temperature of different materials beneath the surface, favoring the ascent of fluids \citep{2020_De_Sanctis_b}.\par
As for the other cases studied in this section, the $S_{480-800nm}$ slope shows the most evident variations. At a larger scale (see Fig. \ref{Fig_SLOPE_480_800}), a system of curved ejecta rays is particularly visible in the southwest-southeast. The farthest ejecta formed a circular to elliptical light violet area around the crater, where the slope is slightly positive to null. Closest to the crater and based on our spectral slope definitions, we identified six areas of different spectral behaviors. The area n$\degr2$, while still on the ejecta, represents the surroundings of the crater and has a value of the $S_{480-800nm}$ around $2.27\times10^{-5}$ \reciprocal{\kilo\angstrom}. The slope decreases on the ejecta blanket, in particular on the west and southwest sides (area and spectrum n$\degr4$) where $S_{480-800nm}$ is about $8.19\times10^{-6}$ \reciprocal{\kilo\angstrom}. The area n$\degr6$, which is defined as a large annular ring (excluding area n$\degr4$), exhibits spectral behavior between the area n$\degr2$ and n$\degr4$. The Occator crater floor (area and spectrum n$\degr5$) has the lowest values of $S_{480-800nm}$, being around $4.78\times10^{-6}$ \reciprocal{\kilo\angstrom}. Area n$\degr$3 corresponds to a part of the rim and the outer ejecta blanket where $S_{480-800nm}$ is a bit higher, reaching a value of $2.17\times10^{-5}$ \reciprocal{\kilo\angstrom}, close to the surrounding terrains' $S_{480-800nm}$ values. Area n$\degr$3 is also slightly visible through the $S_{800-950nm}$ indicator, and it is interesting to note that \cite{2019_Raponi_a} reports a higher abundance of Mg-phyllosilicates and Mg-carbonates in that area.\par
Spectral slopes, $S_{405-465nm}$ and $S_{800-950nm}$ , do not present important variations at this scale on Occator, except on the faculae. The southwest-northeast-oriented line across the Occator crater floor and crossing the Cerealia facula, only visible on the $S_{405-465nm}$ and $S_{800-950nm}$ maps as a low values area, is likely a smearing artifact associated to the high signal acquired on the facula. The crater floor and the closest ejecta (areas n$\degr3$, n$\degr4$ and n$\degr5$) have one of the lowest $S_{405-465nm}$ values of Ceres ($1.60\times10^{-4}$ \reciprocal{\kilo\angstrom} < $S_{405-465nm}$ < $1.64\times10^{-4}$ \reciprocal{\kilo\angstrom}). As illustrated by the map of $I/F_{550nm}$ in Fig. \ref{Fig_IoF_550}, the variation in reflectance is very important at Occator. On Cerealia facula, the $I/F_{550nm}$ reaches a median of 0.19 at \unit{550}{\nano\meter}, with a maximum as high as 0.26 for some observations. For comparison, the $I/F_{550nm}$ of the nearby Vinalia facula is around 0.10 for the brightest observations, while values for Haulani or Oxo reach 0.07. On the contrary, the ejecta located on the northern part have one of the lowest $I/F_{550nm}$, around 0.030 for areas n$\degr2$ and n$\degr3$. At this location \cite{2019_Raponi_a} report  a higher abundance of NH4-phyllosilicates and, while it is not necessarily intuitive, a lower abundance of dark material.\par
On the Occator faculae, the combination between the brightness of the faculae and the integration time leads to an incorrect response of the VIS detector. For this reason, the dataset used here does not allow for a properly study of the faculae. In order to obtain a reliable spectrum, we use a cube (524703945) which is less affected by artifacts acquired during the Ceres Extended Low Altitude Mapping Orbit (CXL) mission phase. Figure \ref{Fig_Occator_zoom} includes the spectrum extracted from this cube (n$\degr1$) and is relative to the Cerealia facula. Vinalia facula is not covered in the cube. The spectrum n$\degr1$ has a different global shape, compared to the other one already discussed. While the $S_{405-465nm}$ is not extreme (around $1.76\times10^{-4}$ \reciprocal{\kilo\angstrom}), this part of the spectrum is straighter, and it extends up to larger wavelengths. On the contrary, while most of the Ceres spectra have a $S_{480-800nm}$ relatively straight after \unit{500}{\nano\meter}, the spectrum of Cerealia facula is much more curved on this range. This does not prevent the $S_{480-800nm}$ to be the highest observed on Ceres, with a value as high as $4.03\times10^{-5}$ \reciprocal{\kilo\angstrom}. This is almost two times the value of the area n$\degr2$ and higher than for Ernutet's organic rich terrain, which reaches $3.39\times10^{-5}$ \reciprocal{\kilo\angstrom}. The slope, $S_{800-950nm}$ , is also high on the faculae, around $-1.19\times10^{-5}$ \reciprocal{\kilo\angstrom} ($-6.47\times10^{-6}$ \reciprocal{\kilo\angstrom} on Ernutet for example).\par
To establish a link between the composition or the state of the surface through the visible spectral indices on the faculae of Occator is not easy. Sodium carbonates are present on the faculae \citep{2016_De_Sanctis, 2018_Carrozzo}, as well as ammonia-bearing and sodium chloride species -- but to a lesser extent (\cite{2016_De_Sanctis, 2016_Ammannito, 2019_Raponi_a, 2020_De_Sanctis_b}). While the spectral shape of the VIS spectrum and its associated slopes and reflectance values are undoubtedly influenced by this peculiar composition, it is not possible to distinguish the separate effect of each component. 
%
\subsection{Oxo}
\label{subSec_Oxo}
\begin{figure*}
    \centering
    \includegraphics[width=17cm]{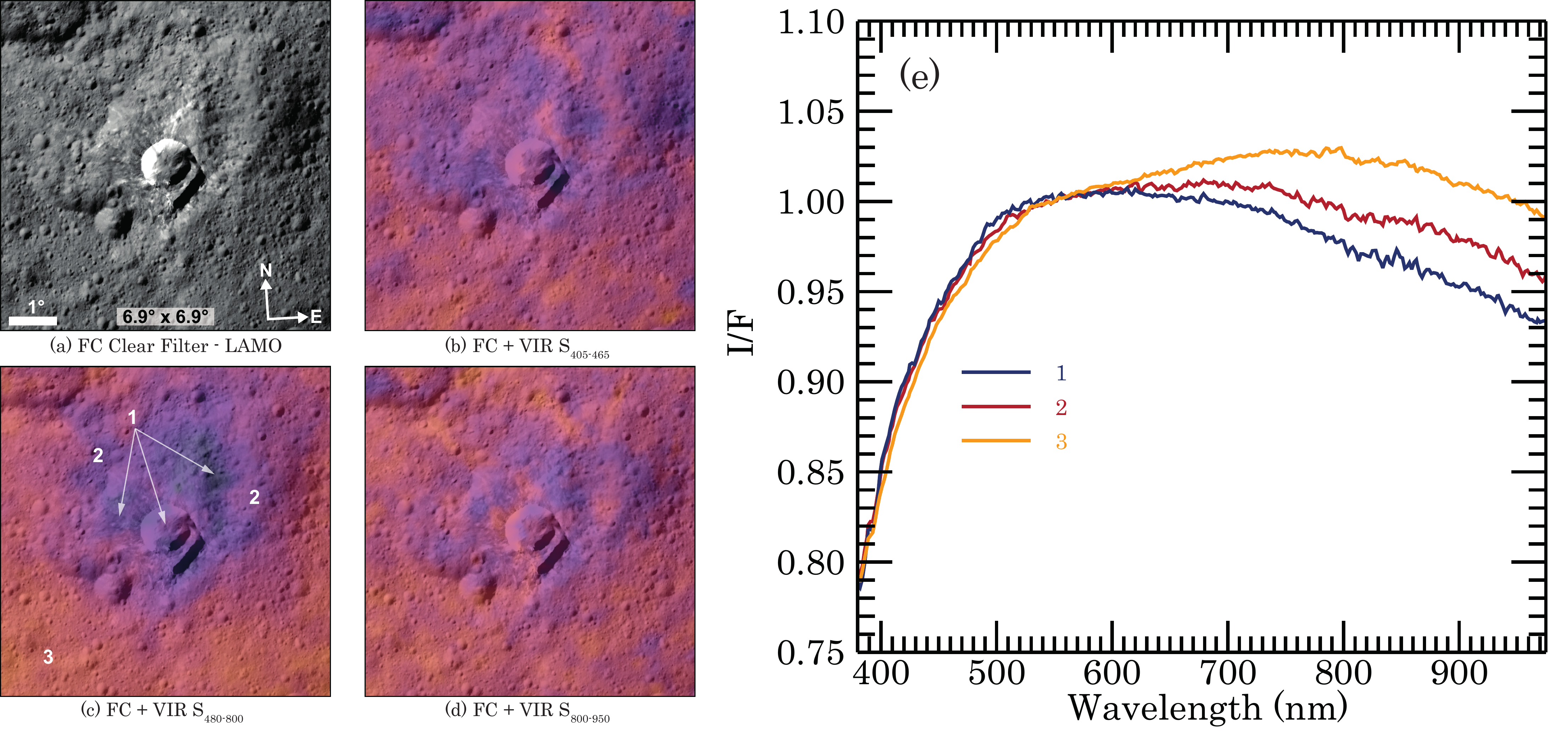}
    \caption{\label{Fig_Oxo_Zoom} Same as Fig. \ref{Fig_Ahuna_Mons_Zoom} for Oxo. Oxo has a diameter of \unit{10}{\kilo\meter}. All the images use a spherical projection with a field of $6.9\degr$ by $6.9\degr$ and have the same orientation.}
\end{figure*}

Oxo is a \unit{10}{km}-diameter, geologically young crater located at $0\degr$E-$45\degr$N. The ejecta of Oxo are poorly extended and among the three spectral slopes -- $S_{405-465nm}$, $S_{480-800nm}$ , and $S_{800-950nm}$ -- only $S_{480-800nm}$ is well correlated with those ejecta, as illustrated in Fig. \ref{Fig_Oxo_Zoom}. The crater floor exhibits a spectral behavior similar to the closest ejecta, and both those areas have been merged in the spectra n$\degr1$ of Fig. \ref{Fig_Oxo_Zoom}. The $S_{480-800nm}$ slope of this area shows one of the lowest values on the surface of Ceres -- reaching $S_{480-800nm}\simeq8,49\times10^{-7}$ \reciprocal{\kilo\angstrom} -- while the surrounding area and spectrum n$\degr3$ are typical of the major part of the Ceres surface, with a value of about $S_{480-800nm}\simeq2.07\times10^{-5}$ \reciprocal{\kilo\angstrom}. A distinct gradation is visible through the $S_{480-800nm}$ slope between the crater floor and the closest ejecta (n$\degr1$ on Fig. \ref{Fig_Oxo_Zoom}), the surrounding terrain (n$\degr3$), and farther ejecta from the crater (spectra and area n$\degr2$).\par
The slopes $S_{405-465nm}$ and $S_{800-950nm}$ on the ejecta exhibit values very close to the surrounding terrains even if some heterogeneities are still visible. However, the level of variation does not allow us to draw any conclusions about the genuineness and significance of the small variations.\par
As seen in the FC image of Fig. \ref{Fig_Oxo_Zoom}, the Oxo crater and its ejecta are particularly bright. Our VIR observations report a reflectance $I/F_{550nm}$ around 0.047 for the area n$\degr1$, while the area n$\degr3$ is around 0.035. As reported by \cite{2016_Combe}, the signature of water ice has been identified on the south rim of Oxo. The VIR observations in the visible area of the same pixels do not allow for a peculiar spectral behavior to be highlighted and they are not reported nor discussed here. \cite{2018_Carrozzo} report a high abundance of carbonates in Oxo. While this high concentration could correlate with a lower $S_{465-800nm}$ spectral slope in the northern ejecta and the crater floor, this is not the case in the southwestern part, where the carbonate concentration is also high.
\section{General discussion}
\label{Sec_General_discussion}
In this section, we focus our interest on the global maps presented in Sect. \ref{Sec_Maps of the spectral parameters}. The spectrum of the sunlight reflected by a surface is the result of a combination of the surface mineralogy along with several physical effects and processes acting on the surface. At visible wavelengths, the Ceres spectrum is devoid of any complete absorption band. Nevertheless, we used several spectral indices -- reflectance, colors, color ratio, and spectral slopes -- to characterize the surface of Ceres and observe the variations of its properties.\par
\subsection{Reflectance and color composites}
\label{SubSec_Discuss_IoF_Color}
The $I/F_{550nm}$ (Fig. \ref{Fig_IoF_550}) map obtained in this study is comparable to the ones discussed by \cite{2017_Ciarniello}, whose photometric correction is also used for producing the $I/F_{550nm}$ map presented in Sect. \ref{SubSec_IoF 550nm map}, and \cite{2018_Longobardo}, which were also based on the VIR data but with a slightly different dataset and without taking advantage of the new slope correction presented in Sect. \ref{SubSec_Data correction and processing}. The VIS and the IR reflectance are relatively similar but some differences may be noted \citep{2017_Ciarniello, 2019_Frigeri}. In \cite{2017_Ciarniello}, an albedo difference map between the $I/F_{550nm}$ and the $I/F_{2\micro\meter}$ was shown and some areas like Haulani, Occator, and Oxo appear brighter in the VIS than in the IR. Among them, Haulani is the most evident, which can be recognized directly in the IR albedo map, being characterized by IR-dark ejecta. For the others, such differences are more subtle. These VIS-IR comparisons also apply to the $I/F_{550nm}$ map presented in the study (Fig. \ref{Fig_IoF_550}). This latter shows very good qualitative agreement with the maps produced by \cite{2016_Nathues, 2017_Schroder} and \cite{2019_Li} based on the Framing Camera observations; small differences may be due to a different approach in the photometric correction and we do not discuss this here. \par
The RGB (Fig. \ref{Fig_RGB}) and the RGB ratio (Fig. \ref{Fig_RGB_ratio}) composites allow for much more information to be highlighted than in the reflectance map, which only reveals only the variation of albedo. The bands of the RGB composites have been chosen to correspond to the ones of \cite{2016_Nathues} and \cite{2017_Schroder}, who have worked with the data of the Framing Camera; thus, the results are very similar, in particular with regard to the maps presented by \cite{2017_Schroder}. The RGB composite map of Sect. \ref{SubSec_Color composite maps}, Fig. \ref{Fig_RGB}, is also comparable to the one presented by \cite{2017_Ciarniello}, but it takes advantage of the correction developed by \cite{2019_Rousseau_c} and is based on a more complete dataset.\par
The first RGB composite of Ceres (Fig. \ref{Fig_RGB}) only allows us to distinguish the most important color and albedo variations. Such albedo differences are visible between the bright Vendimia Planitia and dark Hanami Planum; between the north and the south of Dantu (bright and dark, respectively) or between the northeast of Occator (very dark ejecta) and the Juling and Kupalo region (very bright), for example. The color variability, as appearing in our composite, span between the bluest craters like Haulani, Occator (except the faculae), Ikapati, Centeotl, or Tawals, and the red areas like the Occator facula, the Juling crater floor, the central peak of Urvara and the surroundings of Ernutet while the major part
of the surface is displayed as grey-beige. The Occator Cerealia faculae and the Ernutet surroundings are the reddest places on the Ceres surface. In the latter case, organic-rich material has been identified by \cite{2017_De_Sanctis}. More rarely, green-blue colors are observed, like at Ahuna Mons (which appears greener than blue as mentioned by \cite{2017_Schroder}) or Xevioso. Dantu northern ejecta displays a deeper green color. This deep green is not visible elsewhere, highlighting the dichotomy observed on the Dantu crater and ejecta, as already noted in Fig. \ref{Fig_RGB}. At a local scale, some of the VIS bright features are bluish. However, we do not observe a clear correlation between the albedo and the color at global scale, neither directly in the first RGB composite (Fig. \ref{Fig_RGB}) nor by comparing the $I/F_{550nm}$ and the RGB ratio maps (Figs. \ref{Fig_IoF_550} and \ref{Fig_RGB_ratio}).\par
\subsection{Slopes}
\label{SubSec_Discuss_Slopes}
Spectral slopes are not fully diagnostic of surface composition, as opposed to absorption bands. They instead give clues about a number of factors by which they are influenced, such as: 1) the variation of the composition; 2) the grain size (e.g., \cite{1967_Adams_Filice,1992_Britt}); 3) the structure of the sample and the mixing modalities when several species coexist in a powder or a regolith (e.g., \cite{2011_Cloutis_a, 2012_Cloutis_f,2016_Poch, 2018_Rousseau}); 4) the space weathering effects (e.g., \cite{2004_Moroz, 2005_Nesvorny, 2006_Lazzarin, 2011_Lucey, 2013_Lantz}).\par
These various processes serve as possible explanations for the spectral slope variations observed by VIR in the VIS on the surface of Ceres. It is not the purpose of this study to discuss each one in detail, however, in the following sections, when applicable, we consider whether one or several of these processes can be favored as an explanation of our observations.
The three main spectral slopes that we defined for Ceres in the VIS show marked changes, as shown in the various maps of Sect. \ref{Sec_Maps of the spectral parameters}. We notice that $S_{480-800nm}$ brings out most of the information. The slope $S_{800-950nm}$ highlights almost the same surface features but with less clarity. This is partly explained by the larger wavelength range used to define the $S_{480-800nm}$ slope, which allows for sharpening the variations. However, it does not rule out that particular processes acting on the surface may be preferably visible in that range. For its part, the $S_{800-950nm}$ can be impacted by the same factors -- then toward the VIS wavelengths -- while we cannot exclude that the broad absorption band present around \unit{1.1}{\micro\meter} \citep{2011_Rivkin, 2015_de_Sanctis_b, 2019_Raponi_c} plays also a role if it varies across the surface. As described in the previous sections, the $S_{480-800nm}$ and the $S_{800-950nm}$ slopes mainly highlight  the features related to impact craters and their ejecta. As discussed below, this generally corresponds to the "blue" material as mentioned by \cite{2016_Nathues, 2017_Stephan} and \cite{2017_Schroder}. Because of their peculiar composition, only a couple of exceptions behave differently and show positive slope (or less negative for $S_{800-950nm}$). This is the case for the material present at Ernutet, Urvara's central peak, and the Occator faculae. We also observe several crater floors showing the same positive slopes (Juling, Braciaca, Cacaguat and an unnamed crater north of Haulani ($7.7\degr$E-$20.7\degr$N)), while the others do not. Those crater floors also exhibit a red color in the RGB ratio. This peculiar behavior (also visible as a red color in the RGB ratio map, also noted by \cite{2017_Stephan}) is observed only for few craters and argues for a difference in the composition of the crater floor. \cite{2019_Frigeri} report the variations of the spectral slopes between \unit{1.163}{\micro\meter} and \unit{1.891}{\micro\meter} and between \unit{1.891}{\micro\meter} and \unit{2.250}{\micro\meter}. The $S_{480-800nm}$ and the $S_{800-950nm}$ slopes follow roughly the same trend observed in the IR spectral slopes. This suggests that the factors involved in the variations of the spectral slopes induce the same behavior from the visible to the near-infrared.\par
The $S_{405-465nm}$ slope can also be affected by the same processes that act on the $S_{480-800nm}$ and $S_{800-950nm}$ slopes. However, its behavior is very distinct, and the spectrum of Ceres in this range experiences an important drop toward the UV. This means that other surface properties can be monitored through this spectral indicator. Absorptions due to crystal field effects, conduction bands, (intervalence) charge transfers, and color centers (or F-centers) may occur at these wavelengths (\cite{1977_Hunt, 1982_Sherman, 1989_Burns, 1999_Clark}). However, crystal field effects bands are generally narrow (with respect to charge transfer in \ce{Fe} or \ce{Fe-O}), and these processes are certainly not responsible for the absorption \citep{1977_Hunt, 1999_Clark}. On the other hand, conduction bands, while envisaged by \cite{2016_Hendrix} based on Hubble Space Telescope observations of Ceres, have a sharp edge toward the visible \citep{1973_Johnson_Fanale}, which seems to be incompatible with what we observed with VIR. In addition, they occur in minerals that are neither expected (e.g., sulfur compounds, \cite{1977_Hunt}), nor have been observed on Ceres so far. Finally, absorptions linked to color centers occur in bright minerals (e.g., sulfur, quartz, sodium chloride), but the presence and eventually the quantity needed to be responsible of the absorption observed in the VIR VIS spectra is not compatible with the global reflectance level of Ceres. On the contrary, the band attributed to a charge transfer is common in terrestrial minerals, with its band center is located in the UV with the long-wavelength edge occurring in the VIS; charge transfer absorptions are typically hundreds to thousands of times stronger than the crystal fields bands. Considering that, two processes may cause an absorption compatible with the range of the VIR observations. The first would be the charge-transfer between iron and oxygen, \ce{O2- -> Fe^3+} \citep{1982_Sherman}, which could be linked, on Ceres, to the presence of a Fe-bearing phyllosilicate (e.g., antigorite), or at the presence of the magnetite; all being present or potentially present on the surface \citep{2011_Rivkin, 2015_de_Sanctis_b}. The second would be the metal-to-metal charge-transfer transition, which could be schematized as \ce{2Fe^{3+} -> Fe^{2+} + Fe^{4+}} \citep{1978_Kennedy_Frese}, which causes an absorption feature in the same range. This latter charge-transfer could occur by starting with the oxidation of the serpentine which leads to a degradation of the mineral structure and to an augmentation of the \ce{Fe^3+} in this structure. Finally, the charge transfer, compatible with the drop observed in Ceres' spectrum before \unit{500}{\nano\meter}, may experience variations on the surface. In that case, the deeper the band, the steeper the $S_{405-465nm}$, which then would be a proxy of these processes and consequently of the abundance of the host mineral(s).\par
\subsection{The blue material}
\label{SubSec_Discuss_Bluemat}
While not specific to Ceres \citep{2008_Jaumann}, terrains with negative spectral slopes compared to the Ceres average have been noted by \cite{2016_Nathues, 2017_Schroder} and \cite{2017_Stephan} over the Framing Camera range. The bluish regions on Ceres’ surface are clearly visible in the ratio of images acquired by the FC using filters centered at the wavelengths \unit{438}{\nano\meter} and \unit{749}{\nano\meter}, the F8 and F3 filters, respectively \citep{2017_Stephan}. \cite{2016_Schmedemann} also show the existing age-dependency observed between the blue material and the age of the surface, with the bluest being the youngest. The RGB ratio and the $S_{480-800nm}$ maps (Figs. \ref{Fig_RGB_ratio} and \ref{Fig_SLOPE_480_800}) show a similar spectral behavior than what is observed with the FC. Indeed, a majority of craters and ejecta, as well as few geological features of endogenous origin (e.g., Ahuna Mons) appear bluer than the globally red surface. In parallel to the FC images, \cite{2017_Stephan} made use of the VIR infrared data to confirm the observed blueing. Here, we confirm this observation thanks to the corrected VIR visible data.\par
The origin of the blueing is thoroughly discussed in \cite{2017_Stephan}, who favor a change in particle sizes or an amorphization of the phyllosilicates (with other materials not excluded) to explain this spectral behavior. On the other hand, due to the presence of water-ice in the subsurface of Ceres \citep{2016_Prettyman, 2016_Hiesinger}, \cite{2017_Schroder} favor a process similar to what \cite{2016_Poch} produced in the laboratory, that is, blue spectral behavior of an intra-mixture of smectite and water ice, even after the sublimation of the water ice. In that case, the blueing is due to the foam-like structure of the residue \citep{2016_Poch}. This latter explanation, which has been reproduced with a Ceres-like mixture by \cite{2019_Schroeder} and \cite{2020_Schroeder} is not favored by \cite{2017_Stephan} because of the possible difficulty in keeping the water-ice stable in the ejecta during the impact process and the ejecta deposition. We suggest that the two explanations are not incompatible. The scenario of \cite{2017_Schroder, 2019_Schroeder} could be more likely in the vicinity of the craters. Indeed, lobate flows and knobby crater floors are observed across Ceres \citep{2016_Buczkowski}. The solutions put forward by \cite{2017_Stephan} could play a role on the ejecta, whatever its distance to the crater, explaining why even the farthest exhibits a blue color and spectral slope. The observations of ejecta, like the ones of Haulani or Oxo, where a distinct ejecta blanket is observed close to the crater and is associated with a very blue color and low $S_{480-800nm}$ slope, while the rest of the ejecta are farther and less blue (see Secs. \ref{subSec_Haulani} and \ref{subSec_Oxo}), support this.\par
Over time, the Ceres surface tends to become redder due to the combination of various processes, such as impact gardening, space weathering, and changes in the surface physical properties, such as grain size \citep{2016_Schmedemann, 2017_Stephan, 2017_Schroder}. The current result, as observed by the FC and the VIR in the visible, is a globally red Ceres surface that is punctuated by the blue material coming from the most recent impacts and by some endogenous processes.
\subsection{Comparison with the VIR infrared composition maps}
\label{SubSec_Discuss_VISvsIR}
The global mapping of Ceres in the infrared revealed a widespread abundance of phyllosilicates \citep{2015_de_Sanctis_b, 2016_Ammannito} and \ce{Mg}-carbonates with localized spots of \ce{Na}-carbonates (\cite{2016_De_Sanctis, 2018_Carrozzo}; see also \cite{2019_McCord} for further references).
\subsubsection{Phyllosilicates}
\label{SubsubSec_Phyllosilicates}
The \ce{Mg}- and \ce{NH_4}- phyllosilicate maps of \cite{2016_Ammannito} have a similar general trend but present small differences. We observe that the variations of the $S_{480-800nm}$ qualitatively follow one of the \ce{Mg}-phyllosilicate abundances: where the abundances of the \ce{Mg}-phyllosilicate are lower (e.g., Haulani, Ikapati, south Dantu, Occator, Juling, and Kupalo), the $S_{480-800nm}$ slope is also lower. Some exceptions can be noted, such as for Ernutet and Urvara.\par
Contrary to $S_{480-800nm}$, $S_{405-465nm}$ is more similar to the \ce{NH_4}-phyllosilicate map. In particular, within the Vendimia Planitia region and close to the Urvara central peak, $S_{405-465nm}$ is steeper and the \ce{NH_4}-phyllosilicates are more abundant, following \cite{2016_Ammannito}. However, the opposite behavior is observed for the crater centered at $138\degr$E--$24\degr$S: $S_{405-465nm}$ is higher but \cite{2019_De_Sanctis} report a lower abundance of phyllosilicates.\par
\subsubsection{Carbonates}
\label{SubsubSec_Carbonates}
Carbonates are globally present on Ceres as \ce{Mg}-carbonates and in localized spots associated with impact craters, fractures, cryovolcanic structures, or bright spots as \ce{Na}-carbonates \citep{2016_De_Sanctis, 2017_Zambon, 2018_Carrozzo}. The global distribution of \ce{Mg}-carbonates, which is very homogeneous, does not correlate with the behavior of the VIS spectrum through the various spectral parameters investigated in Sect. \ref{Sec_Maps of the spectral parameters}.\par
Concerning \ce{Na}-carbonates, a high abundance is observed in Occator faculae \citep{2016_De_Sanctis} and Urvara (eastern rim and central peak), matching a larger $S_{405-465nm}$. However, this trend is not observed in Ahuna Mons (Fig. \ref{Fig_Ahuna_Mons_Zoom}), Ikapati, Oxo (Fig. \ref{Fig_Oxo_Zoom}), Azacca, Ernutet (Fig. \ref{Fig_Ernutet_Zoom}), or Haulani (Fig. \ref{Fig_Haulani_Zoom}), where \ce{Na}-carbonates are also safely identified \citep{2018_Carrozzo}. On their side, the $S_{480-800nm}$ and the $S_{800-950nm}$ slopes show bluer behavior when \ce{Na}-carbonates are present (except for Ernutet and the Urvara central peak), but an equal trend is observed in very different locations, as well as when \ce{Na}-carbonates are absent. Thus, no direct correlations exist between spectral slopes and carbonates.\par
Regarding the RGB ratio composite, we may observe a distinct green-blue color in certain areas where \ce{Na}-carbonates are detected. This is the case for Ahuna Mons, Xevioso, an unnamed crater in the northeast of Xevioso (located at $318\degr$E-$8.7\degr$N and designed as "crater 1" by \cite{2018_Carrozzo}), and, to a lesser extent, for Kupalo. However, the trend is not always the same: for example, Kupalo is rich in \ce{Na}-carbonates but only mildly green-blue in the RGB ratio composite. On the contrary, the Cacaguat crater appears green-blue in the RGB ratio composite but no \ce{Na}-carbonates seems to be detected on the map presented by \cite{2018_Carrozzo}. Since the VIS color is driven by different processes, other locations with a green-blue color in Fig. \ref{Fig_RGB_ratio} and associated with already existing carbonates could be masked because of the dominance of the blue color -- corresponding to the fresh material excavated from the crater -- or because of a lower abundance of carbonates or, finally, because of various surface properties. Thus, this correlation is not obvious and is not observed for all \ce{Na}-carbonates-rich areas. \cite{2018_Tosi} and \cite{2018_Carrozzo} show that a process of dehydration could happen for various hydrated \ce{Na}-carbonate species.It may be speculative but if a link exists between the green-blue color and the carbonates, its origin could possibly be found through this process.\par
\section{Conclusion}
\label{Sec_Conclusion}
Using the newly corrected VIR VIS dataset of Ceres \citep{2019_Rousseau_c}, we characterized its surface in the visible by means of various spectral parameter maps. Our maps have been made available for the community through the Aladin Desktop software. Our spectral parameters highlight marked and distinct changes at both global and local scales, testifying to variations in the composition and in the physical properties of the Ceres' surface.\par
We show that the main driver of the changes in the color and of the $S_{480-800nm}$ spectral slope (and to a lesser extent, $S_{800-950nm}$) are the impact craters, which tend to turn the crater floors and the ejecta blue. This has already been noticed in Framing Camera observations \citep{2016_Nathues, 2017_Stephan, 2017_Schroder}. We observe some exceptions, such as north of Dantu and various red crater floors (e.g., Juling), indicating that differences in the surface properties may occur there. Features of endogenous origin may be bluer (e.g., Ahuna Mons) or redder (Occator faculae) than the global Ceres surface. As already observed, the organic material at Ernutet appears red through the $S_{480-800nm}$ and the $S_{800-950nm}$ spectral slopes, but is not outlined by the $S_{405-465nm}$.\par
The $S_{405-465nm}$ spectral slope characterizes the drop of the Ceres spectrum observed toward the UV. We suggest that the \ce{O2- -> Fe3+} and/or the \ce{2Fe^{3+} -> Fe^{2+} + Fe^{4+}} charge transfer may be responsible for this absorption and that the $S_{405-465nm}$ slope could be a proxy to follow its variations. However, no mineral phases can be identified on the unique basis of the visible data. The $S_{405-465nm}$ slope presents other behaviors compared to $S_{480-800nm}$ and $S_{800-950nm}$ slopes, evidencing different surface features (crater and ejecta are less visible compared to $S_{480-800nm}$ slope). In particular, the north of Dantu is well highlighted as in the RGB composite maps. The $S_{405-465nm}$ and RGB ratio characteristics at Dantu therefore suggest that the Dantu region may differ, in terms of composition and surface physical properties, compared to the rest of the surface of Ceres.\par
At visible wavelengths, beyond \unit{550}{\nano\meter}, the Ceres visible spectrum is devoid of absorption bands. This implies that no mineral species with a signature in the visible are identified. If present, they exist in a small amount and the dark phase of the surface may mask them easily. We compared the abundance maps of the phyllosilicates and carbonates with the visible parameters of this study. We observed some correlations between the \ce{Mg}-phyllosilicates and the slope, $S_{480-800nm}$, and between the \ce{NH_4}-phyllosilicates and the slope, $S_{405-465nm}$. In the latter case, the correlation is particularly strong in the Dantu region. Concerning the carbonates, no correlation exists between visible spectral parameters and the \ce{Mg}-carbonates. We observe a partial correlation between the distribution of the \ce{Na}-carbonates and the VIS spectral slopes and color. However, the variations of the VIS slopes cannot be attributed to their presence alone.
\begin{acknowledgements}
VIR is funded by the Italian Space Agency (ASI) and was developed under the leadership of INAF-Istituto di Astrofisica e Planetologia Spaziali, Rome, Italy (Grant ASI INAF I/004/12/0). The instrument was built by Selex-Galileo, Florence, Italy. The authors acknowledge the support of the Dawn Science, Instrument, and Operations Teams.
The authors made use of TOPCAT (Tools for OPerations on Catalogues And Tables, \cite{2005_Taylor}) for a part of the data analysis and figure production.
This research has made use of "Aladin Desktop" developed at CDS, Strasbourg Observatory, France \citep{2000_Bonnarel, 2015_Fernique}. We thank Sharon Uy (UCLA, USA), who greatly helped for manuscript editing. We thank the anonymous reviewer for the insightful suggestions which improved the paper.
\end{acknowledgements}
\bibliographystyle{aa.bst}
\bibliography{Bib.bib}
\begin{appendix}
\section{Density maps}
\label{Appendix_Density_maps}
\begin{figure*}[h]
    \centering
    \includegraphics[width=17cm]{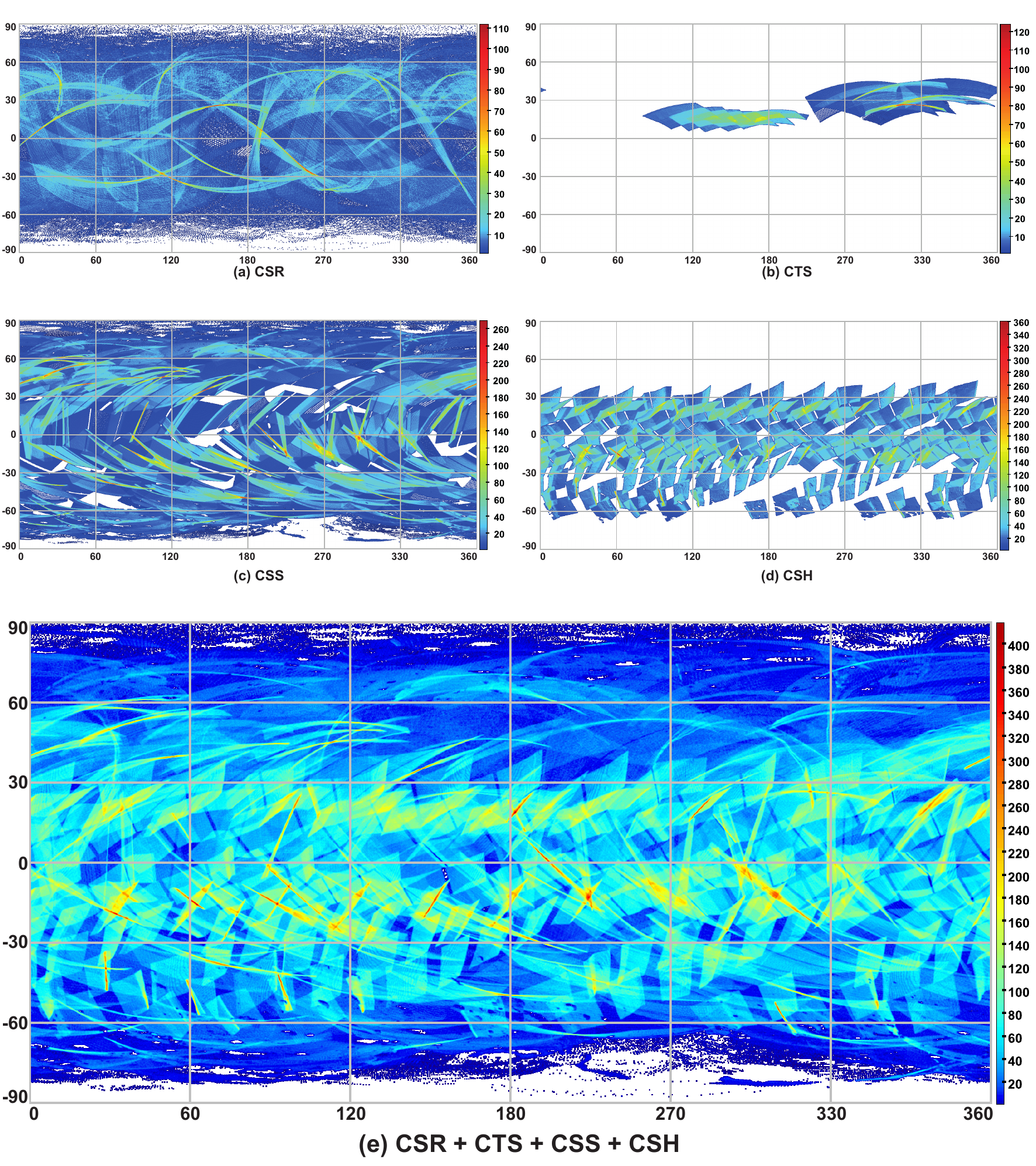}
    \label{Fig_Appendix_DENSITY_MAP}
    \caption{Density maps of the VIR visible data set used in the study. Panel A corresponds to the CSR mission phase; panel B to the CTS; panel C to CSS; panel D to CSH; and panel E regroups every four. For details about the mission phases, see Table \ref{Table1} and Sect. \ref{SubSec_Data correction and processing}. Each map is built with TOPCAT with a Plate Carée projection (see Sect. \ref{SubSec_Map projections}), and observations are represented as points. The scale corresponds to the square root of the observation density.}
\end{figure*}
\section{Framing Camera LAMO map and main features}
\label{Appendix_LAMO_names_map}
\begin{figure*}[h]
    \centering
    \includegraphics[width=17cm]{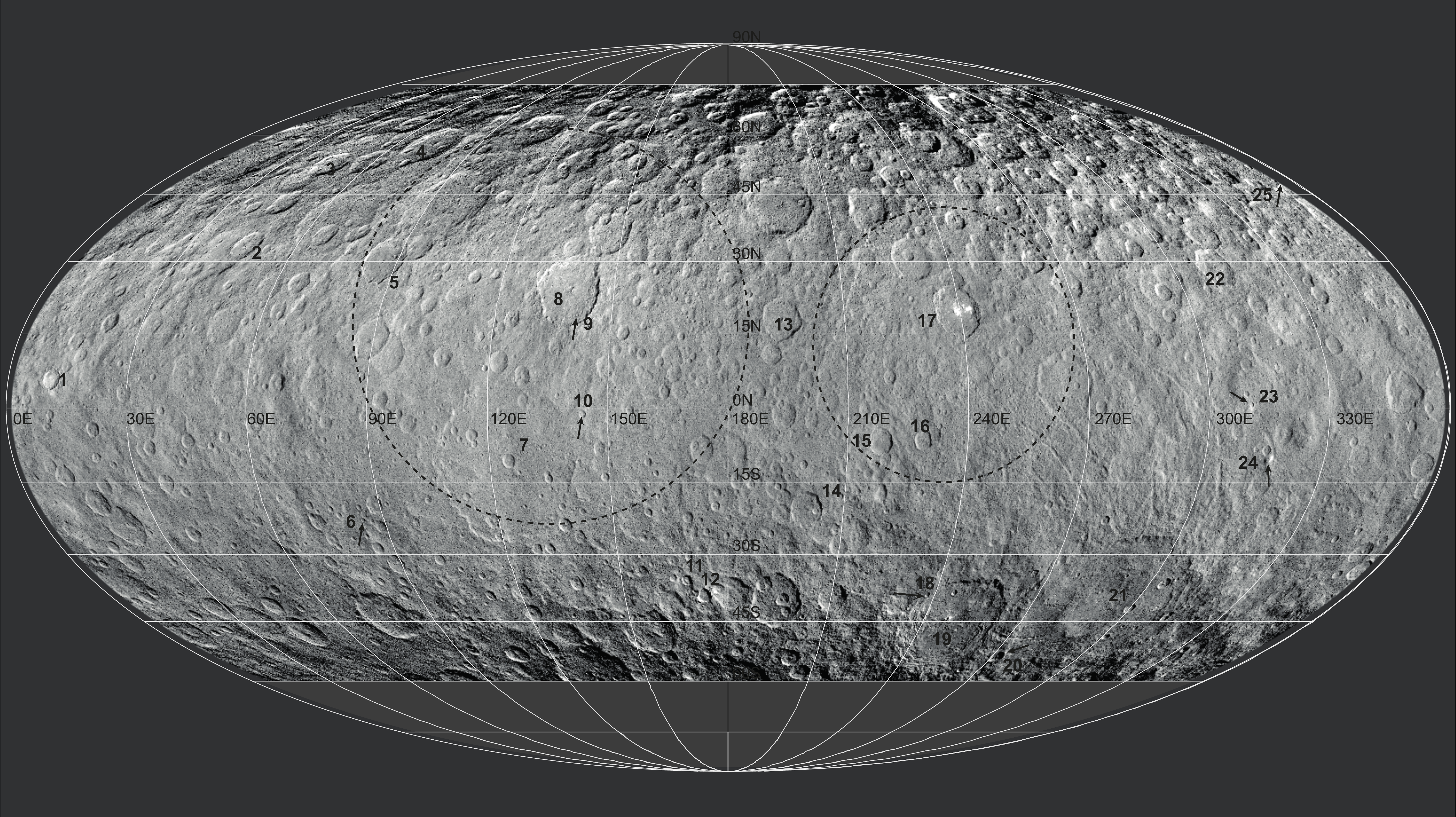}
    \label{Fig_Appendix_LAMO_MAP}
    \caption{Framing Camera LAMO map from \cite{2016_Roatsch_b} reprocessed as a HiPS and with a Mollweide projection. The map is used as background context for the maps of the spectral slopes in Figs. \ref{Fig_SLOPE_405_465}, \ref{Fig_SLOPE_480_800} and \ref{Fig_SLOPE_800_950}, as well as for the maps of Sect. \ref{Sec_Area_of_interest}. The dashed ellipse on the left correspond broadly to Vendimia Planitia, to the right of Hanami Planum. The main numbered features are: 1) Haulani; 2) Ikapati; 3) Ernutet; 4) Omonga; 5) Gaue; 6) Braciaca; 7) Kerwan; 8) Dantu; 9) Centeotl; 10) Cacaguat; 11) Juling; 12) Kupalo; 13) Nawish; 14) Consus; 15) Azacca; 16) Lociyo; 17) Occator; 18) Tawals; 19) Urvara; 20) Nunghui; 21) Yalode; 22) Fejokoo; 23) Xevioso; 24) Ahuna Mons; and 25) Oxo}.
\end{figure*}
\section{Maps of the spectral slopes without Framing Camera context and coordinate grid}
\begin{figure*}
    \centering
    \includegraphics[width=17cm]{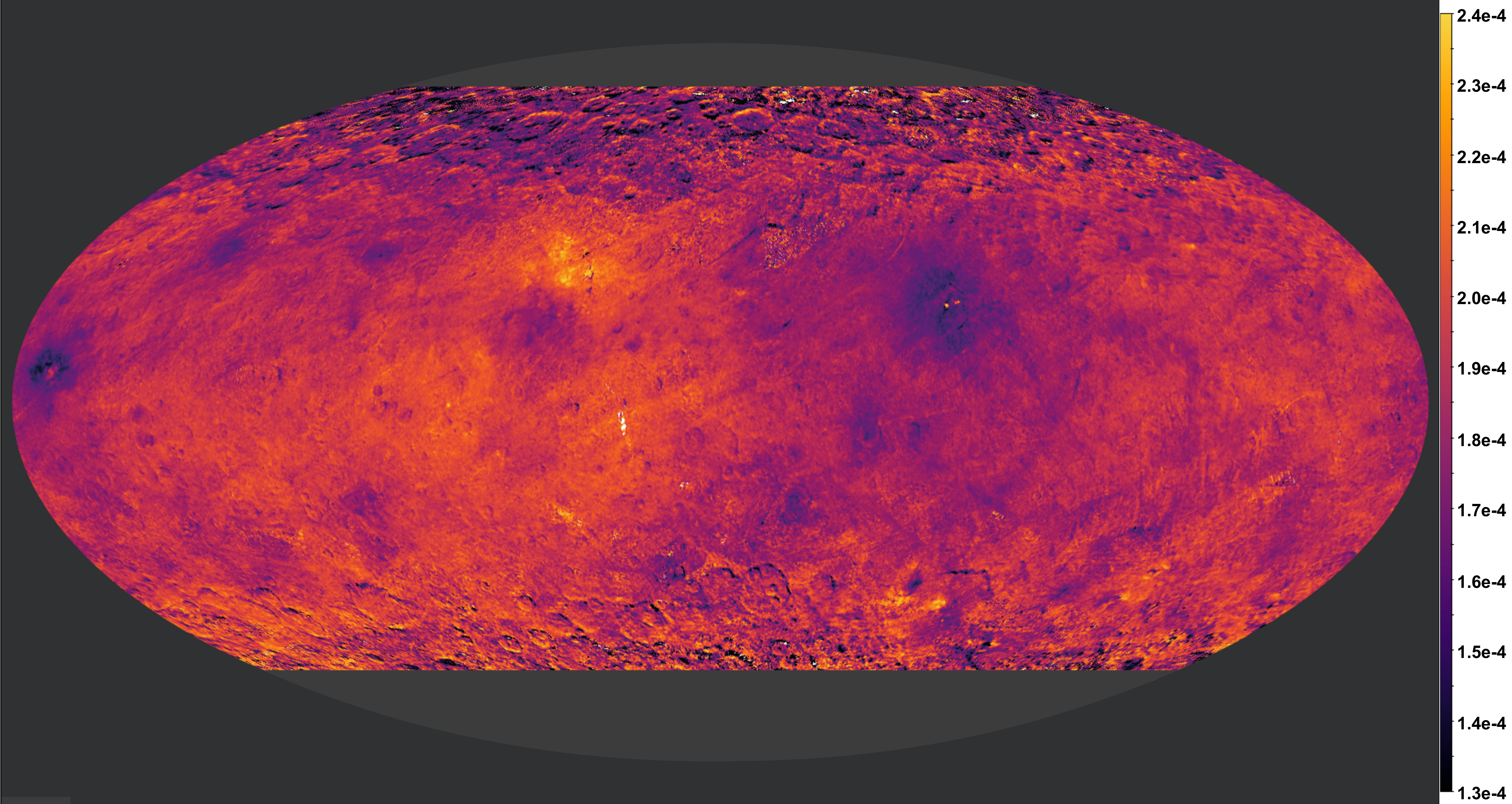}
    \caption{\label{Fig_APPENDIX_SLOPE_405_465_NoTransp_NoGrid} Map of the VIR $S_{405-480nm}$ spectral slope without transparency effect and Framing camera context. White areas correspond to missing data.}
\end{figure*}
\begin{figure*}
    \centering
    \includegraphics[width=17cm]{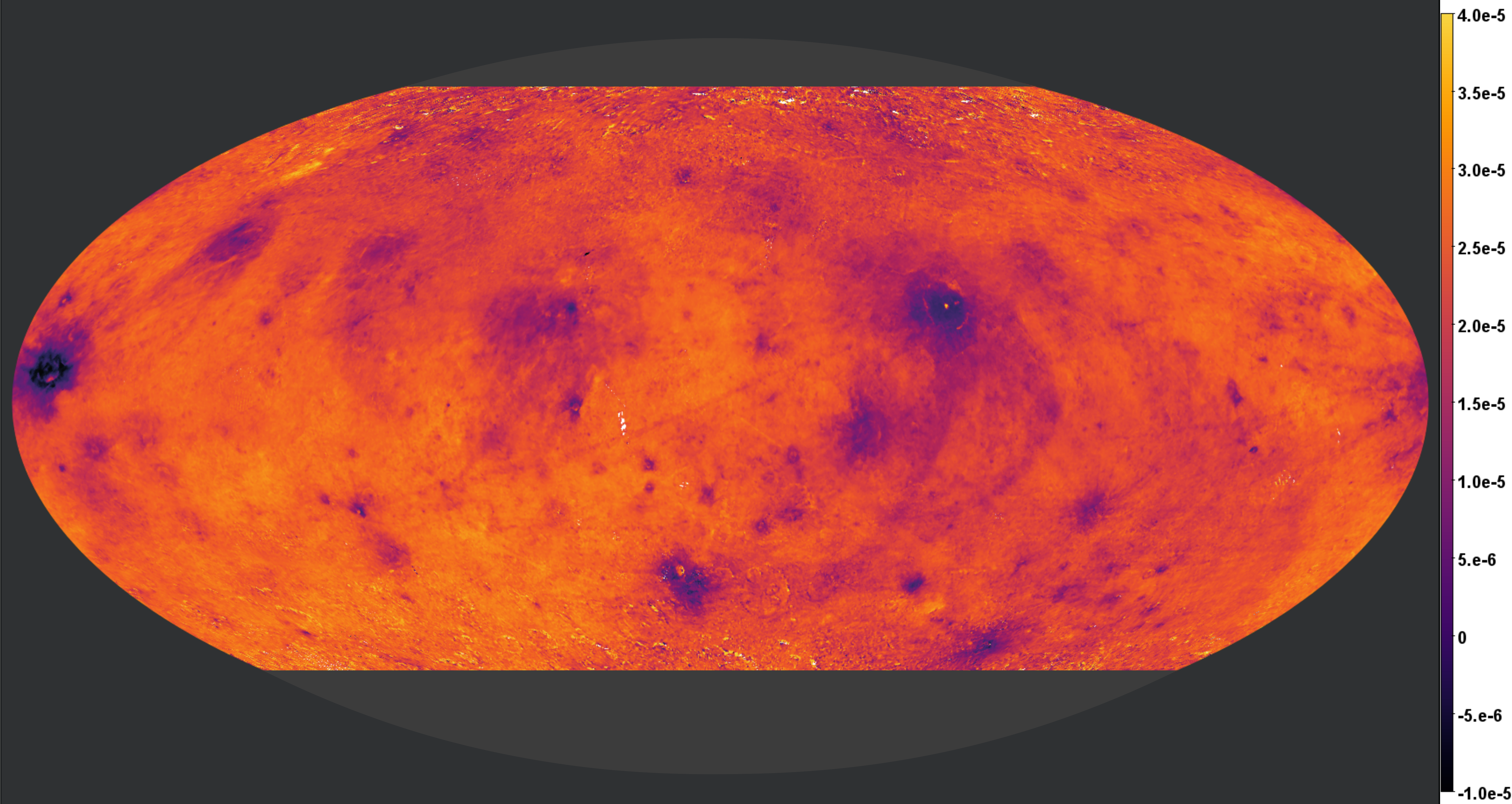}
    \caption{\label{Fig_APPENDIX_SLOPE_480_800_NoTransp_NoGrid} Map of the VIR $S_{480-800nm}$ spectral slope without transparency effect and Framing camera context. White areas correspond to missing data.}
\end{figure*}
\begin{figure*}
    \centering
    \includegraphics[width=17cm]{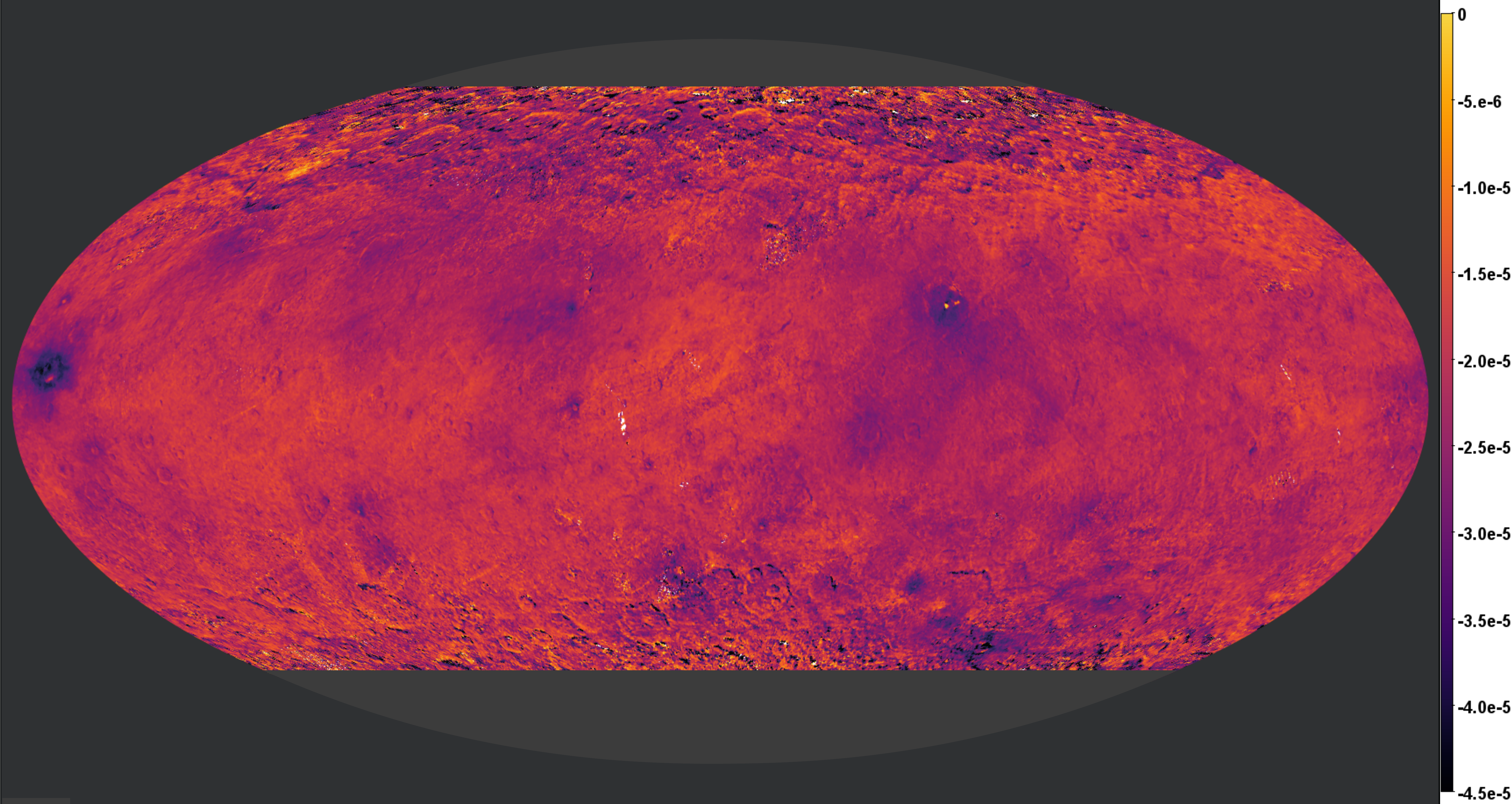}
    \caption{\label{Fig_APPENDIX_SLOPE_800_950_NoTransp_NoGrid} Map of the VIR $S_{800-950nm}$ spectral slope without transparency effect and Framing camera context. White areas correspond to missing data.}
\end{figure*}
\end{appendix}
\end{document}